\pdfoutput=1
\documentclass[letterpaper,preprint,fleqn,prd,superscriptaddress,nofootinbib,longbibliography]{revtex4-2}

\usepackage{amsfonts,amssymb,amsmath,mathtools,mathrsfs,slashed}	
\usepackage{graphicx}		
\usepackage{hyperref}		
\usepackage{array}
\usepackage[utf8]{inputenc}
\usepackage[shortlabels]{enumitem}
\usepackage{physics}
\usepackage{color}
\usepackage{comment}
\usepackage{dsfont}
\usepackage{soul}
\usepackage{xcolor}

\def\be{\begin{equation}}
\def\ee{\end{equation}}
\def\ba{\begin{eqnarray}}
\def\ea{\end{eqnarray}}
\newcommand{\pr}[1]{\left(#1\right)}
\newcommand{\pq}[1]{\left[#1\right]}

\def\nb{\nonumber}
\def\p{\partial}

\def\a{A}

\def\t{\tau}

\begin{document}
\begin{flushright}
CPHT-RR051.062024 \\
HIP-2024-17/TH
\end{flushright}

\title{Spinodal slowing down and scaling in a holographic model}

\author{Alessio Caddeo}
\email{caddeoalessio@uniovi.es}
\affiliation{Department of Physics and \\
Instituto de Ciencias y Tecnolog\'{\i}as Espaciales de Asturias (ICTEA),\\
Universidad de Oviedo, c/ Leopoldo Calvo Sotelo 18, ES-33007 Oviedo, Spain}

\author{Oscar Henriksson}
\email{oscar.henriksson@abo.fi}
\affiliation{Faculty of Science and Engineering, {\AA}bo Akademi University, Henrikinkatu 2, 20500 Turku, Finland}
\affiliation{Helsinki Institute of Physics, P.O.~Box 64, 00014 University of Helsinki, Finland}

\author{Carlos~Hoyos}
\email{hoyoscarlos@uniovi.es}
\affiliation{Department of Physics and \\
Instituto de Ciencias y Tecnolog\'{\i}as Espaciales de Asturias (ICTEA),\\
Universidad de Oviedo, c/ Leopoldo Calvo Sotelo 18, ES-33007 Oviedo, Spain}

\author{Mikel Sanchez-Garitaonandia}
\email{mikel.sanchez@polytechnique.edu}
\affiliation{CPHT, CNRS, \'Ecole polytechnique, Institut Polytechnique de Paris, 91120 Palaiseau, France}

\begin{abstract}
The dynamics of first-order phase transitions in strongly coupled systems are relevant in a variety of systems, from heavy ion collisions to the early universe. Holographic theories can be used to model these systems, with fluctuations usually suppressed. In this case the system can come close to a spinodal point where theory and experiments indicate that the the behaviour should be similar to a critical point of a second-order phase transition. We study this question using a simple holographic model and confirm that there is critical slowing down and scaling behaviour close to the spinodal point, with precise quantitative estimates. In addition, we determine the start of the scaling regime for the breakdown of quasistatic evolution when the temperature of a thermal bath is slowly decreased across the transition. We also extend the analysis to the dynamics of second-order phase transitions and strong crossovers.
\end{abstract}

\maketitle

\newpage

\tableofcontents

\section{Introduction}

Matter going through a first-order phase transition as it cools down is a commonly observed occurrence in daily life, as when we turn water into ice cubes for our drinks. At much smaller subatomic scales, extremely hot matter produced in heavy ion collisions may undergo a transition as a droplet of quark-gluon plasma expands and cools down, turning into hadronic matter \cite{Pandav:2022xxx,Du:2024wjm}. In the other extreme, at cosmological scales, an electroweak or dark matter phase transition in the early universe may leave an observable imprint in the gravitational wave spectrum \cite{Hindmarsh:2020hop,Athron:2023xlk}. If observed, the former would validate phenomenological predictions for the QCD phase diagram, while the latter would be a clear signal of physics beyond the Standard Model.

Holographic models have been used to describe and estimate the effects of such phase transitions, for both heavy ion collisions \cite{Critelli:2018osu,Attems:2018gou,Rougemont:2022piu} and cosmology \cite{Ares:2020lbt,Zhu:2021vkj,Ares:2021nap,Bea:2021zol,Escriva:2022yaf,Chen:2022cgj,Bigazzi:2020phm,Bigazzi:2020avc}. Such models use a classical gravitational theory to compute non-trivial quantities of a dual strongly coupled field theory. In particular, one can describe the out-of-equilibrium evolution of the field theory as it goes through a phase transition by solving a few classical differential equations. This makes the holographic duality a powerful tool to study various aspects of such transitions, including the nucleation of bubbles and the properties of domain walls \cite{Bigazzi:2021ucw,Bea:2021zsu,Janik:2021jbq,Henriksson:2021zei,Ares:2021ntv,Bea:2021ieq,Janik:2022wsx,Bea:2022mfb,Wang:2023lam,Li:2023xto,Sanchez-Garitaonandia:2023zqz}, the formation and evolution of mixed phases \cite{Attems:2017ezz,Janik:2017ykj,Attems:2019yqn,Bellantuono:2019wbn,Yang:2019ibe,Bea:2020ees,Attems:2020qkg,Zhao:2023ffs}, and quenches \cite{Bhaseen:2012gg,Basu:2013soa,Garcia-Garcia:2013rha,Sonner:2014tca,Das:2014lda,Chesler:2014gya,Ishii:2015gia,Natsuume:2017jmu,Xia:2021pap,Flory:2022uzp,Chen:2022cwi,Chen:2022tfy,Xia:2024wfq}.

In this paper, we continue this program by using a simple holographic model to study slow evolution through first-order phase transitions, comparing the results to second-order transitions and crossovers. 

Holographic duality is typically used in the classical gravity limit, which, in the dual field theory, is a large-$N$ or mean field limit. In the canonical case $N$ is the rank of the gauge group \cite{Aharony:1999ti}, but more generally we can think of it as a measure of the number of degrees of freedom. Restricting to a classical gravitational description is therefore akin to completely removing fluctuations around the mean field theory. Even if $N$ is not strictly infinite, in order for the holographic approach to be tractable, fluctuations must be strongly suppressed. This can have a significant effect on the evolution of the system through a phase transition.

\begin{figure}[t!]
\begin{center}
\includegraphics[width=0.62\textwidth]{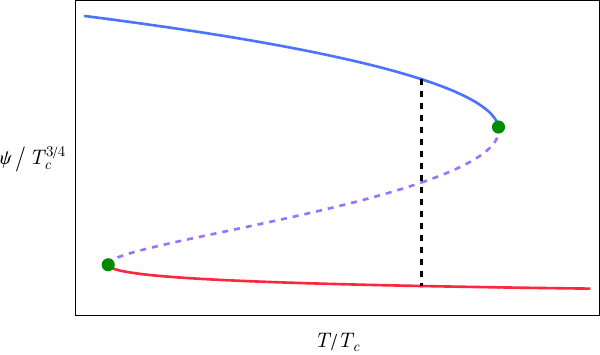}
\end{center}
\caption{Order parameter as a function of the temperature for the case in the first row of Table \ref{tab:parametVeff}. The blue and red lines are the (meta)stable low and high temperature phases respectively. The dashed purple line is the unstable phase. The dashed black vertical line marks the critical temperature, and the green dots the position of the spinodal points where metastable and unstable phases merge.}
\label{fig:FOPTvev}
\end{figure}

Consider the case of a system with a first-order phase transition. The order parameter as a function of the temperature would typically behave as shown in Figure~\ref{fig:FOPTvev}. Within some interval of temperatures around the critical value, the high- and low-temperature phases coexist; the one with lower free energy is the true stable phase, while the other is metastable. Precisely at the critical temperature the free energy of the two phases is the same; the phase which is stable at larger temperatures is metastable at smaller temperatures, and vice versa. Each of the stable phases ends at a \emph{spinodal point}, marked as green dots in the figure, where they merge with an unstable third branch.\footnote{Note that this is a mean field, or coarse grained, depiction of the phase diagram, since the true free energy must be a convex function of thermodynamical variables. The true equilibrium states include mixtures of the two phases.} 

We will study the case where the system of interest is put in contact with a large thermal bath that determines the temperature, which we then change with time. If the change is slow enough, the system could in principle remain arbitrarily close to equilibrium, undergoing a quasistatic evolution of equilibrium quantities.

Starting at high temperatures and cooling down, the system will typically undercool, reaching temperatures below the critical one before fluctuations drive it to the low temperature phase. Usually, the fluctuations take the form of bubbles of the low temperature phase nucleating inside the high temperature phase, which expand due to their higher inner pressure (lower free energy).  If the nucleating barrier is very low, the size of the bubble can become of the same order of the width of the interface and in this case one should also take into account the formation of ramified domains of the low temperature phase \cite{Binder1987}. As argued in \cite{Ares:2021ntv,Bea:2021zol}, bubble nucleation is exponentially suppressed with $N$ in holographic models, so, for a large enough value of $N$, the time evolution can be slow enough as to remain quasistatic, but fast enough to completely neglect nucleation before reaching the vicinity of the spinodal point. In other words, the large-$N$ limit can uncover the physics of the spinodal points, which would otherwise be hidden by the nucleation process.

Although not widely discussed, spinodal points share some features with critical points of second-order phase transitions, for example the scaling behaviour of thermodynamic variables and correlators, as well as the {\em critical slowing down} of perturbations \cite{Compagner1974,Fisher1982,Sasaki2007}. These features may be relevant even in the presence of thermal fluctuations and disorder, as long as they are not too strong \cite{Liang2015,An:2017brc,Zhong2017,Banerjee2023}, and they have been observed experimentally in some strongly correlated materials \cite{Bar2018,Zhu2018,Furukawa2018,Kundu_2020,Pal_2020,Bar2021,Kumbhakar2022,Kundu2023}. The critical slowing down means that the time it takes for an out-of-equilibrium perturbation to decay diverges as one approaches the critical or spinodal points. Thus, even if the evolution is very slow, eventually the system is driven out of equilibrium before reaching the critical or spinodal temperature. 

In this paper we provide a simple theoretical realization of a strongly correlated system where the properties of transitions through spinodal and critical points can be studied, confirming the above picture. By tuning the parameters of our holographic model, we can realize phase transitions of first and second order, as well as smooth crossovers. In the case of first-order transitions, we show that the system can never reach the spinodal point; no matter how slow the time evolution is, the system will fall out of equilibrium before reaching it, even in the absence of bubble nucleation. 
We will also show evidence for the scaling behaviour at the spinodal point in the holographic model and compare with scaling at critical points and strong crossovers.  Although our results are obtained for a particular class of models, we expect the qualitative features to be valid more generally for phase transitions in strongly coupled theories with a holographic dual.

The rest of the paper is organized as follows: In section \ref{sec:holographicModel} we introduce the holographic model we work with, and discuss its equilibrium configurations and phase diagram. In section \ref{sec:fluct}, we introduce small perturbations around equilibrium and extract the relaxation time and correlation length. We argue analytically, and show numerically, that the model exhibits critical slowing down at the spinodal points. In section \ref{sec:breakdown}, we determine the breakdown of quasistatic evolution and the scaling close to critical and spinodal points, both through analytic approximations and numerically. We conclude and discuss possible future directions in section \ref{sec:discussion}. Several appendices go into more detail regarding the phase diagram, the computation of decay rates and correlation lengths, quasinormal modes, and the numerical methods used.

\section{The holographic model and its equilibrium behavior}\label{sec:holographicModel}

We will work with a bottom-up gravitational model dual to a $2+1$-dimensional strongly coupled CFT with a large number $\sim N_c^2$ of degrees of freedom. Conformal invariance will be broken in a subsector involving $\sim N_c$ degrees of freedom, akin to fundamental fields spanning the flavor sector of a gauge theory with a gauge group of rank $N_c$. We will work in the quenched approximation where the effect of this `flavor' subsector over the rest of the theory is neglected. 

The gravity dual to this setup consists of Einstein gravity with a cosmological constant plus a scalar field. The classical action is
\begin{equation}\label{eq:action}
    S = \frac{1}{2\kappa_4 ^2} \int d^4 x \sqrt{|g|}\left[ R+\frac{6}{L^2}  -\frac{1}{N_c}\left(g^{MN} \partial_M \phi \partial_N \phi + 2P(\phi)\right) \right]\ ,
\end{equation}
where $g_{MN}$ is the metric, $R$ the associated scalar curvature and $\phi$ the scalar field with a potential $P(\phi)$. The metric $g_{MN}$ corresponds to the `glue' sector with $\sim N_c^2$ degrees of freedom, while the scalar $\phi$ will capture the physics of the flavor sector. Following the quenched approximation, we will treat the scalar as a probe, neglecting the backreaction over the metric. The metric at zero temperature will then be AdS$_4$ with radius $L$.

When turning on the temperature the glue sector will act as a thermal bath for the flavor sector. Since there are no phase transitions in the glue sector, for a time evolution slow enough (compared to the scale set by the temperature), we can assume that the evolution is quasistatic, i.e. the system is at thermal equilibrium at each instant of time. We will then fix the metric to be AdS$_4$ black brane with a position of the horizon $z_h$ that may depend on time
\begin{equation}\label{eq:adsBHmetric}
    ds^2 = \frac{L^2}{z^2} \left(  f(z) dt^2 + \frac{dz^2}{f(z)} + dx^i dx^i\right) \ , \quad \qquad f(z) = 1- \frac{z^3}{z_h ^3} \ .
\end{equation}
The temperature in the dual field theory equals the Hawking temperature of the black brane geometry,
\begin{equation}\label{eq:temperature}
    T=\frac{3}{4\pi z_h} \ .
\end{equation}
If $z_h$ depends on time this metric is not a solution of Einstein's equations, but can serve as a good approximation as long as the the time derivatives are small enough. 

The flavor sector will be at the temperature set by the bath. Normally it should also follow a quasistatic evolution as the temperature is changed, but when it comes close to a phase transition we expect this will no longer be true. In order to elucidate the evolution through the phase transition we will solve the equations of motion of the dual scalar field in the background metric with a time-dependent horizon. We will assume homogeneous configurations in the spatial directions, so there will be dependence only in time and the holographic radial direction. However, before this can be done, it will be necessary to set up the model such that it shows a phase transition. This can only happen if conformal invariance is broken in some way.

\subsection{Breaking of conformal invariance}

We will now specify the potential for the scalar field. For small values of the amplitude it typically starts with a quadratic term
\begin{equation}
    P(\phi)= \frac{m^2}{2} \phi^2+\cdots \ .
\end{equation}
The asymptotic boundary expansion of the scalar field in the AdS$_4$ metric is determined by the value of the mass in units of the AdS radius. One finds two independent solutions: 
\begin{equation}
    \phi(z)\sim \phi_- z^{\frac{3}{2}-\nu} +\phi_+ z^{\frac{3}{2}+\nu} \ ,\qquad \nu=\sqrt{m^2L^2+\frac{9}{4}} \ ,
\end{equation}
with $\phi_-$ and $\phi_+$ the coefficients of the leading and subleading solutions. For $\nu\geq 3/2$ the leading solution maps to a source, or coupling, for the dual operator and the subleading solution determines the expectation value. However, for $3/2>\nu>0$ one may choose an `alternative quantization' where the roles of the two solutions are reversed \cite{Breitenlohner:1982bm,Klebanov:1999tb}. In the case of ordinary quantization the conformal dimension of the dual operator is $\Delta_+=\frac{3}{2}+\nu$, while for alternative quantization it is $\Delta_-=\frac{3}{2}-\nu$.

The simplest way to break conformal invariance is to turn on a source for a relevant operator, which will amount to solving the equations of motion for the scalar field fixing $\phi_-$ or $\phi_+$. It can be shown that this is not enough to induce a phase transition with only a quadratic potential. A possible solution is to design a more complicated potential $P(\phi)$, for instance by adding higher powers of the scalar field with appropriate coefficients as in \cite{Bea:2018whf}. We will instead follow the approach of \cite{Ares:2021ntv}, where the bulk potential $P(\phi)$ remains quadratic, but a non-trivial effective potential is obtained by using the alternative quantization for the scalar field and tuning its boundary conditions. This corresponds to deforming the dual field theory by various powers of the operator $\mathcal{O}$ dual to $\phi$: $S_{CFT}\to S_{CFT} + g_n \mathcal{O}^n$ \cite{Witten:2001ua}. In a large-$N$ theory such a deformation simply results in the addition of a similar term to the effective potential depending on the expectation value of the operator, $V(\langle \mathcal{O}\rangle)\to V(\langle \mathcal{O}\rangle) + g_n \langle \mathcal{O}\rangle^n $ \cite{Papadimitriou:2007sj}.

We want to construct an effective potential which is quartic in $\langle \mathcal{O}\rangle$, as this will let us realize first- and second-order thermal phase transitions, as well as smooth crossovers. In order not to have irrelevant deformations of the CFT, this means that the conformal dimension of the operator should be $\Delta \leq 3/4$. We find it convenient to saturate this condition, meaning that $m^2L^2=-27/16$. Implementing both alternative quantization and the boundary conditions requires adding additional boundary terms to the gravitational on-shell action, after it has been properly renormalized, as we now review.

\subsection{Holographic renormalization and alternative quantization}

Since we are working in the probe approximation, the Einstein and scalar actions are renormalized independently. For us only the scalar action is relevant. Following the usual procedure of holographic renormalization, we introduce a cutoff in the holographic radial coordinate $z=z_{UV}$, and add a boundary counterterm of the form
\begin{equation}
    S_{c.t.}=-\frac{3}{8L\kappa_4^2N_c}\int d^3 x \sqrt{|h|}\phi^2\Big|_{z=z_{UV}} \ ,
\end{equation}
where $h_{\mu\nu}=g_{\mu\nu}\Big|_{z=z_{UV}}$ is the induced metric on a radial slice.

The variation of the regularized scalar on-shell action is
\begin{equation}
    \delta S_\phi =\delta S_{\text{on-shell}}+\delta S_{c.t.} = \frac{1}{\kappa_4 ^2N_c} \int d^3 x \left(\sqrt{|g|} \, g^{zz}\partial_z \phi -\frac{3}{4L}\sqrt{|h|}\phi\right)\delta\phi \Big|_{z=z_{UV}}\ .
\end{equation}
Taking the $z_{UV}\to 0$ limit, we obtain the variation of the generating functional of the dual field theory in ordinary quantization
\begin{equation}
    \delta {\cal W}_{\Delta_+}[\phi_-]=\frac{3L^2}{2\kappa_4 ^2N_c}\int d^3 x \,\phi_+\delta \phi_- \ .
\end{equation}
In order to move to alternative quantization we add the boundary term \cite{Papadimitriou:2007sj}
\begin{equation}
    S_{a.q.}=-\frac{3L^2}{2\kappa_4^2N_c}\int d^3x\, \phi_+\phi_- \ .
\end{equation}
Combining this with the variation of the renormalized on-shell action gives the variation of the generating functional of the dual field theory when the operator has dimension $\Delta_-=3/4$:
\begin{equation}\label{eq:vargf}
    \delta {\cal W}_{\Delta_-}[\phi_+]= \delta S_\phi\Big|_{z_{UV}\to 0}+\delta S_{a.q.}=-\frac{3L^2}{2\kappa_4 ^2N_c}\int d^3 x \,\phi_-\delta \phi_+ \ .
\end{equation}
The generating functional itself (rather than the variation) can also be explicitly computed in this case. First, the on-shell action for the scalar is 
\begin{equation}
    S_\phi=\frac{1}{2\kappa_4^2 N_c}\int d^3 x\,\sqrt{|g|} g^{zz}\partial_z\phi \,\phi\,\Big|_{z=z_{UV}}^{z=z_h} \ .
\end{equation}
For regular solutions the contribution at the horizon vanishes. Then, adding all the boundary terms, the generating functional is
\begin{equation}
    {\cal W}_{\Delta_-}[\phi_+]=-\frac{3L^2}{4\kappa_4 ^2N_c}\int d^3 x \,\phi_+\phi_- \ .
\end{equation}

\subsection{Effective potential and multitrace boundary conditions}
\label{subsec:effpotentialandmultitracebc}

The effective potential can be obtained from the generating functional via a Legendre transform that trades the dependence on the source by a dependence on the expectation value of the operator. First, we split the coefficient of the subleading term $\phi_+$ into the coupling breaking conformal invariance, $\Lambda$, plus an external source, $J$, i.e. $\phi_+=-\Lambda+J$. The generating functional is
\begin{equation}
    {\cal W}_{\Delta_-}[\Lambda,J]=-\frac{3L^2}{4\kappa_4 ^2N_c}\int d^3 x \,(J-\Lambda)\phi_-[J] \ .
\end{equation}
The expectation value of $\mathcal{O}$ is obtained by varying the generating functional with respect to the source:
\begin{equation}
\label{eq::expvaluepsi}
    \left\langle\cal O\right\rangle =\frac{\delta {\cal W}_{\Delta_-}}{\delta J}=-\frac{3L^2}{2\kappa_4 ^2N_c}\phi_-\equiv \frac{L^2}{2\kappa_4 ^2N_c} \psi \ .
\end{equation}
The Legendre transform of the generating functional results in the effective action
\begin{equation}
    \Gamma_{\Delta_-}[\psi]={\cal W}_{\Delta_-}[\Lambda,J]-J \left\langle\cal O\right\rangle=-\frac{L^2}{4\kappa_4 ^2N_c}\int d^3 x \,(J[\psi]+\Lambda)\psi \ .
\end{equation}
Using that $J=\phi_++\Lambda$, up to an overall factor the effective potential is
\begin{equation}
    V(\psi)=\frac{1}{2}\phi_+[\psi]\psi+\Lambda \psi \ .
\end{equation}
For a quadratic scalar action like ours, the equations of motion for the scalar field are linear, implying that $\phi_+$ will be linear in $\psi$. The effective potential cannot have several minima in this case, so there cannot be a phase transition. 

We can modify the effective potential by adding additional boundary terms, that we write using a function $W(\psi)$ whose meaning will be clear momentarily:
\begin{equation}
    S_W=\frac{L^2}{2\kappa_4 ^2N_c} \int d^3 x\left(\psi \, \partial_\psi W(\psi)- W(\psi)\right) \ .
\end{equation}
With the new terms, the variation of the renormalized on-shell action in alternative quantization becomes
\begin{equation}
    \delta {\cal W}_{\Delta_-}[\phi_+]+\delta S_W=\frac{L^2}{2\kappa_4^2 N_c}\int d^3 x\, \psi \,\delta \left(\phi_++\partial_\psi W(\psi)\right)\equiv \frac{L^2}{2\kappa_4^2 N_c}\int d^3 x\, \psi \,\delta J \ .
\end{equation}
We see that the new terms modify how the source should be identified, thus implementing the multitrace boundary conditions \cite{Witten:2001ua,Papadimitriou:2007sj} where
\begin{equation}\label{eq:J}
     J=\phi_++\Lambda+\partial_\psi W(\psi) \ .
\end{equation}
After performing the Legendre transform to get the effective action, one finds that the effective potential is now modified to
\begin{equation}\label{eq:effpot}
     V(\psi)=\frac{1}{2}\phi_+[\psi]\psi+\Lambda \psi+W(\psi) \ .
\end{equation}
Note that for a linear relation between the coefficients of leading and subleading terms $\phi_+[\psi]\propto \psi$ \footnote{This will be true only in the static case, otherwise $\phi_+$ also depends on time derivatives of $\psi$.}, the source is equal to the first derivative of the potential
\begin{equation}
    J=\partial_\psi V(\psi) \ .
\end{equation}
Thus, configurations with no source coincide with the critical points of the effective potential. We will have to impose the condition $J=0$ in order to stay within a theory with a fixed coupling $\Lambda$.

Having derived the effective potential \eqref{eq:effpot}, we will pick the multitrace deformation to be a quartic polynomial
\begin{equation}\label{eq:multitrace}
    W(\psi)=\frac{a}{2}\psi^2+\frac{b}{3}\psi^3+\frac{c}{4}\psi^4  \ .
\end{equation}
 We can then tune the parameters $a,b,c,\Lambda$ such that, for example, the effective potential has several critical points and the theory exhibits a phase transition as the temperature changes.

\subsection{Static solutions and phase diagram}
\label{subsec:phasediagram}

To fully specify the effective potential, we now determine the function $\phi_+[\psi]$. This is done by finding the static solutions for the scalar field $\phi=\phi(z)$ in the background AdS$_4$ black brane \eqref{eq:adsBHmetric}. The solutions should satisfy the no-source boundary condition and be regular at the black brane horizon. 

From the scalar action \eqref{eq:action} with a quadratic potential, the equations of motion are
\begin{equation}
    \partial_z\left( \frac{f(z)}{z^2} \partial_z \phi\right)-\frac{m^2L^2}{z^4}\phi=0 \ .
\end{equation}
Taking into account that $m^2L^2=-27/16$, there are two independent solutions that we identify according to the asymptotic expansion at the AdS boundary
\begin{equation}
    \phi(z)=\phi_- z_h^{3/4}\eta_{3/4}\left(\frac{z}{z_h}\right)+\phi_+ z_h^{9/4}\eta_{9/4}\left(\frac{z}{z_h}\right) \ ,
\end{equation}
where
\begin{equation}\label{eq:scalarsols}
   \eta_{3/4}(Z)=Z^{3/4}\,{}_2F_1\left(\frac{1}{4},\frac{1}{4},\frac{1}{2},Z^3 \right) \ ,\qquad \eta_{9/4}(Z)=Z^{9/4}\,{}_2F_1\left(\frac{3}{4},\frac{3}{4},\frac{3}{2},Z^3 \right) \ .
\end{equation}
Expanding close to the horizon, we find
\begin{equation}
    \phi(z)\sim C_h z_h^{-9/4}\left( \phi_++\gamma z_h^{-3/2}\phi_-\right)\log(z_h-z)+\cdots \ ,
\end{equation}
where
\begin{equation}\label{eq:gamma}
    C_h=-\frac{4 \pi ^{3/2}}{\Gamma \left(\frac{1}{4}\right)^2} \ ,\qquad \gamma=\frac{2 \Gamma \left(\frac{3}{4}\right)^2}{\Gamma \left(\frac{1}{4}\right)^2 } \ .
\end{equation}
Regularity of the solution requires the coefficient of the logarithmic term to vanish. The regular solution is the linear combination
\begin{equation}\label{eq:regularsol}
\eta_R(Z)=\eta_{3/4}(Z)-\gamma \, \eta_{9/4}(Z)  \ .
\end{equation}
This fixes the relation between $\phi_+$ and $\phi_-$ in equilibrium solutions
\begin{equation}
    \phi_+=-\gamma z_h^{-3/2}\phi_-=\frac{\gamma}{3} z_h^{-3/2}\psi \ .
\end{equation}
Combining this result with the expression for the temperature \eqref{eq:temperature} and the effective potential \eqref{eq:effpot}, \eqref{eq:multitrace} we arrive at
\begin{equation}\label{eq:effpotT}
     V(\psi)=\Lambda \psi+\frac{1}{2}a_T\psi^2+\frac{b}{3}\psi^3+\frac{c}{4}\psi^4 \ ,
\end{equation}
where the total coefficient of the quadratic term is
\begin{equation}\label{eq:aT}
    a_T=a+\frac{\gamma}{3}\left(\frac{4\pi T}{3}\right)^{3/2} \ .
\end{equation}
Finally, the value of $\psi$ is fixed by the no-source condition $J=\partial_\psi V(\psi)=0$: 
\begin{equation}\label{eq:criticalcond}
  \Lambda+ a_T\psi+b\psi^2+c\psi^3=0 \ .
\end{equation}

The real solutions of \eqref{eq:criticalcond} determine the stationary points of the effective potential \eqref{eq:effpotT} and thus the equilibrium states of the dual field theory. Minima correspond to stable or metastable equilibrium states, while a maximum corresponds to an unstable equilibrium state in the dual field theory. The theory features a first-order phase transition if the potential has a single minimum at high and low temperatures and three stationary points in some intermediate range $T_1>T>T_2$. In appendix~\ref{app:criticalpoints} we analyze in detail the conditions the parameters in the effective potential need to satisfy in order to have a first-order transition, a second-order transition, or a strong crossover. In particular, we define a parameter $\a$ which distinguishes between theories with first-order transitions ($\a>0$), second-order transitions ($\a=0$), and crossovers ($\a<0$). By tuning $\a$ we can then study the interplay between the scaling associated to the second-order transition and to the first-order spinodal point, as well as their effect on strong crossovers near the second-order critical point.

\begin{figure}[t!]
\begin{center}
{\includegraphics[scale=1]{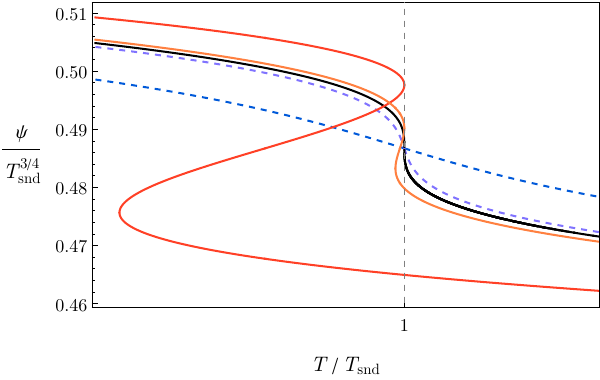}
}
\end{center}
\caption{Equilibrium expectation value of the scalar operator as a function of the temperature. The curves represent: first-order phase transition with $\a=5\cdot 10^{-4}$ (red), $\a=5\cdot 10^{-5}$ (orange); 
second-order phase transition (black); crossover with $\a=- 5 \cdot 10^{-4}$ (blue), and crossover with $\a=- 5 \cdot 10^{-5}$ (purple). Here,
$T_{\text{snd}} \approx 5.162 M$ is the critical temperature of the second-order phase transition.}
\label{fig:EVvsTcurves}
\end{figure}

We will focus on the set of parameters shown in Table~\ref{tab:parametVeff}, for which we present numerical results in section~\ref{sec:numericalresults}. In Figure~\ref{fig:EVvsTcurves}, we display the order parameter as a function of temperature for some of these parameters. In Table~\ref{tab:parametVeff} and elsewhere, we write the parameters in units of a reference mass scale $M$ that can be chosen arbitrarily as part of the definition of the theory.

\begin{table}[h]
    \centering
    \begin{tabular}{|c||c|c|c|c|c|c|c|c|c|}
    \hline
        Order & $\a$ & $\Lambda M^{-9/4}$ &  $a M^{-3/2}$ & $b M^{-3/4}$ &  \ $c$ \ & $T_c M^{-1}$ & $T_1/T_c$ & $T_2/T_c$ & Fig.~\ref{fig:EVvsTcurves}     \\ \hline
         1st & $0.2014$ &  $-1$ & $-1$ & $-5$ & $1$ & $4.934$ & $1.046$ & $0.8135$ & - \\ \hline
         2nd & 0 & $-4.630$ & $0.6765$ & $-5$ & $1$ & $5.162$ & - & - & black\\
         \hline
         $\infty$ & $-5\times 10^{-3}$ & $-4.699$ & $0.7181$ & $-5$ & $1$ & $\sim 5.16$ & - & - & - \\
         \hline
         $\infty$ & $-5\times 10^{-4}$ & $-4.637$ & $0.6806$ & $-5$ & $1$ & $\sim 5.16$ & - & - & blue \\
         \hline
                  $\infty$ & $-5\times 10^{-5}$ & $-4.630$ & $0.6768$ & $-5$ & $1$ & $\sim 5.16$ & - & - & purple \\
         \hline
         1st & $+5 \times 10^{-3}$ & $-4.557$ & $0.6348$ & $-5$ & $1$ & $5.1609$ & $1+2 \cdot 10^{-4}$ & $1-2 \cdot 10^{-4}$ & - \\
         \hline
         1st & $+5\times 10^{-4}$ & $-4.623$ & $0.6723$ & $-5$ & $1$ & $5.1618$  & $1+5 \cdot 10^{-6}$ & $1-6 \cdot 10^{-6}$ & red\\
         \hline         
         1st & $+5 \times 10^{-5}$ & $-4.629$ & $0.6761$ & $-5$ & $1$ & 5.1618 & $1+2 \cdot 10^{-7}$  & $1-2 \cdot 10^{-7}$  & orange \\ \hline
    \end{tabular}
    \caption{Parameters of the effective potential and critical and spinodal temperatures for the phase transitions and crossovers studied in this paper. Some of these entries are also displayed in Figure~\ref{fig:EVvsTcurves}. For the crossovers $T_c$ refers to the approximate temperature where the expectation value of the scalar operator changes rapidly.}
    \label{tab:parametVeff}
\end{table}

\section{Relaxation time and correlation length}\label{sec:fluct}

In this section, we fix the temperature and study small perturbations around thermal equilibrium, in order to show that critical slowing down is a feature of both spinodal points and second-order critical points in our model. We will also obtain the dynamical exponent $z$ relating the correlation length $\xi$ with the relaxation time $\tau$, $\xi\sim \tau^{1/z}$ by considering fluctuations with non-zero spatial momentum.

Starting from the action (\ref{eq:action}) and the background metric (\ref{eq:adsBHmetric}), we first put the metric in Eddington-Finkelstein form with dimensionless coordinates. In order to do this, we rescale the coordinates by the horizon position,
\begin{equation}
z=z_h Z \ , \quad \quad  t=z_h X^0 \ ,\quad  \quad  x^i= z_h X^i\ .
\end{equation}
After this, we perform the change of variables
\begin{equation}
X^0=v+r(Z) \ , \quad  \quad  r'(Z)=\frac{1}{F(Z)}\ , \quad  \quad F(Z)=1-Z^3\ .
\end{equation}
The solution for $r$ is
\begin{equation}
r(Z)=r_0+\frac{1}{6} \left[\log \left(Z^2+Z+1\right)-2 \log (1-Z)+2 \sqrt{3} \tan ^{-1}\left(\frac{2
   Z+1}{\sqrt{3}}\right)\right]\ .
\end{equation}
The resulting metric is
\begin{equation}
ds^2=\frac{L^2}{Z^2}\left(-F(Z)dv^2-2dv dZ+ dX^i dX^i\right)\ .
\end{equation}
The equation of motion we need to solve becomes
\begin{equation}
\label{eq:scalareom}
0= \partial_Z^2 \phi+\left(\frac{F'(Z)}{F(Z)}-\frac{2}{Z} \right)\partial_Z\phi+\frac{27/16}{Z^2 F(Z)}\phi +\frac{2}{Z F(Z)}\partial_v \phi-\frac{2}{F(Z)} \partial_Z\partial_v \phi+\frac{1}{F(Z)}\partial_i^2 \phi\ .
\end{equation}
According to the discussion in section \ref{subsec:effpotentialandmultitracebc}, we need to solve (\ref{eq:scalareom}) imposing regularity at the horizon $Z=1$ and the no-source, multitrace, boundary condition $J=0$, with $J$ given by (\ref{eq:J}).

Consider the regular static solution for the scalar field $\eta_R(Z)$ found in \eqref{eq:regularsol}.  The equilibrium solution satisfying regularity is then
\begin{equation}\label{eq:staticsol}
\phi_0(Z)=-\frac{\overline{\psi}_0}{3}\eta_R(Z),\ \ \psi_0=\left(\frac{4\pi T}{3}\right)^{3/4}\overline{\psi}_0 \ ,
\end{equation}
where $\overline{\psi}_0$ is a constant that should be chosen to satisfy the no-source condition $J=\partial_\psi V(\psi_0)=0$.

We now introduce a small perturbation around this equilibrium solution:
\begin{equation}
\phi(v,Z)=\phi_0(Z)+\delta \phi(v,Z,X) \ .
\end{equation}
Inserting this in \eqref{eq:scalareom}, the equation of motion for the perturbation can be written as
\begin{equation}\label{eq:pertscalareom}
0=\partial_Z \left( \frac{F(Z)}{Z^2}\partial_Z \delta\phi\right)+\frac{27/16}{Z^4}\delta\phi +\frac{2}{Z^3}\partial_v \delta\phi-\frac{2}{Z^2} \partial_Z\partial_v \delta\phi+\frac{1}{Z^2}\partial_i^2 \delta\phi \ .
\end{equation}
We can identify the leading and subleading coefficients in the boundary expansion of the perturbation, 
\begin{equation}
    \delta \phi(v,Z,X)\sim \delta\phi_-(v,X) Z^{3/4}+\delta \phi_+(v,X) Z^{9/4}+\cdots\,. \qquad \delta\phi_-=-\frac{1}{3}\delta \psi \ .
\end{equation}
The no-source boundary condition expanded to linear order determines the boundary condition for the perturbation:
\begin{equation}\label{eq:deltaJ}
\delta J=\delta\phi_+(v,X) +\partial_\psi^2 W(\psi_0)\delta \psi(v,X)=0 \ .
\end{equation}

\subsection{Relaxation time and critical slowing down}
\label{subsec::criticalslowingdown}

We can restrict the analysis to homogeneous perturbations $\delta\phi(v,Z,X)=\delta\phi(v,Z)$. The no-source condition \eqref{eq:deltaJ} and the equation of motion \eqref{eq:pertscalareom} are consistent with an exponential decay of the perturbation:
\begin{equation}
\delta \phi(Z,v)=e^{-\Gamma v} \delta \phi_0(Z) \ ,\qquad \delta\psi(v)=e^{-\Gamma v}\delta\psi_0 \ .
\end{equation}
In appendix \ref{app:decayrate}, we obtain the following analytical estimate of the dimensionless decay rate $\Gamma$, under the assumption that it is small:
\begin{equation}\label{eq:ratio}
\Gamma\approx \frac{9}{2(1-2I_1)}\left(\frac{4\pi T}{3} \right)^{-3/2}V''(\psi_0) \ ,
\end{equation}
where $I_1 \approx 0.141115$. Restoring units, the relaxation time is the inverse of the decay rate
\begin{equation}
\tau^{-1}=\Gamma_T= \frac{\Gamma}{z_h}=\left(\frac{4\pi T}{3} \right)\Gamma\approx \frac{9}{2(1-2I_1)}\left(\frac{4\pi T}{3} \right)^{-1/2}V''(\psi_0) \ .
\end{equation}
If we approach the critical point of a second-order phase transition or the spinodal point of a first-order phase transition, the curvature of the effective potential vanishes, $V''(\psi_0)\to 0$. Then, we have now shown that the decay rate also vanishes $\Gamma_T\to 0$, so that perturbations take longer and longer times to decay. This corresponds to the anticipated critical slowing down.

The decay rate $\Gamma_T$ also corresponds to the purely imaginary frequency $\omega=-i\Gamma_T$ of the lowest quasinormal mode (QNM) in the spectrum of perturbations of the scalar field in the black brane geometry. We have computed numerically the first few QNM as detailed in Appendix~\ref{app:quasinormalmodes}. In Figure \ref{plotcomparegamma}, we show that the numerical results agree with the analytic formula in the vicinity of spinodal points.

\begin{figure}[t!]
\begin{center}
\includegraphics[width=0.48\textwidth]{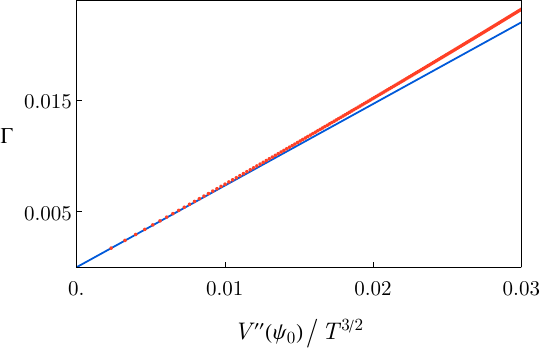}
\end{center}
\caption{Comparison between the (dimensionless) decay rate computed from the analytical formula (\ref{eq:ratio}) (red) and the numerical result obtained as described in appendix \ref{app:quasinormalmodes} (blue) for temperatures close to the spinodal temperature $T_{2}$.}
\label{plotcomparegamma}
\end{figure}

\subsection{Correlation length and dynamical exponent}
\label{subsec::dynamicalexp}

In order to compute the correlation length and dynamical exponent, we expand the perturbation in plane waves,
\begin{equation}\label{eq:planewave}
    \delta\phi(v,Z,X)=\int \frac{d\bar\omega d^3 \bar k}{(2\pi)^3}\delta\phi_{\omega,k}(Z)e^{-i\bar\omega v+i \bar k\cdot X} \ ,
\end{equation}
where $\bar \omega=3\omega/(4\pi T)$ and $\bar k=3 k/(4\pi T)$. 

The two-point function for the order parameter has a pole corresponding to the lowest QNM; we find that the associated dispersion relation reads, for small momenta, 
\begin{equation}\label{eq::k-dependence-dispersion}
    \omega = -i\Gamma_T - i D_T k^2 + \cdots \ ,
\end{equation}
where $D_T$ depends only on $T$, and $\Gamma_T \approx \tau^{-1}$ corresponds to the decay rate previously introduced and shown in Figure \ref{plotcomparegamma}. At zero frequency, the Fourier transform of the two-point function has the form
\begin{equation}
    \left\langle\psi(k)\psi(-k)\right\rangle\approx \frac{Z}{k^2+\xi^{-2}}   \ ,
\end{equation}
where the pole at imaginary momentum $k =  \pm i\xi^{-1}$ determines the correlation length, implying that
\begin{equation}
    \xi^{-2}\approx \frac{\Gamma_T}{D_T} \ .
\end{equation}
Close to the critical and spinodal points the correlation length diverges and, as shown in appendix~\ref{app:decayrate}, we obtain the follwing approximate analytic expression:
\begin{equation}
    \xi^{-2}\approx \frac{9}{2I_2}\left( \frac{4\pi T}{3}\right)^{1/2}V''(\psi_0)\sim \Gamma\sim \tau^{-1} \ ,\qquad I_2\approx 1.829 \ .
\end{equation}
This shows that the exponent  of the momentum in the dispersion relation coincides with the dynamical exponent $z=2$. Thus, it can be obtained by measuring the small momentum dependence of the lowest QNM frequency.
We also obtain the dimensionless ratio
\begin{equation}\label{eq:Dval}
   D\equiv \frac{4\pi T}{3} D_T =\frac{4\pi T}{3}\frac{\xi^2}{\tau}\approx\frac{I_2}{1-2I_1}\approx 2.548 \ .
\end{equation}

\begin{figure}[t!]
\begin{center}
{\includegraphics[width=0.48\textwidth]{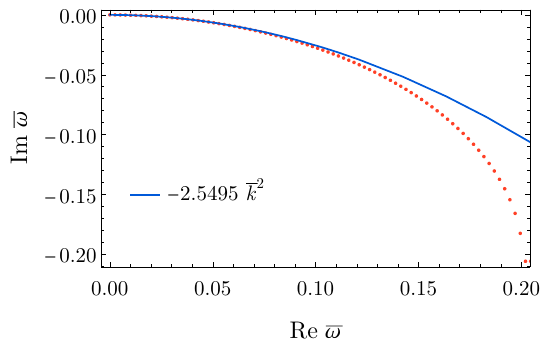}
\includegraphics[width=0.48\textwidth]{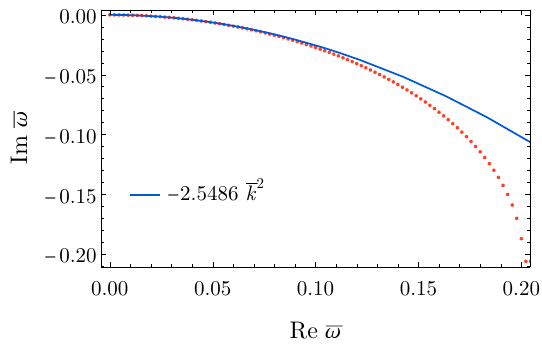}}
\end{center}
\caption{Momentum dependence of the lightest QNM frequency for (on the left) the first-order transition in the first row of Table \ref{tab:parametVeff} near the spinodal temperature, and (on the right) the second-order transition in the second row of Table \ref{tab:parametVeff} near the critical temperature.}
\label{fig::dynamicalExponent}
\end{figure}

Our numerical results are shown in Figure \ref{fig::dynamicalExponent}, where we show the momentum dependence of the frequency of the lightest QNM frequency for the first- and second-order phase transitions in the first two lines of Table \ref{tab:parametVeff}. We have added a $k^2$ curve (gray line), showing that the dynamical exponent is $z = 2$ for both kind of transitions. The numerical calculation also confirms that the dimensionless coefficient $D$ extracted from \eqref{eq::k-dependence-dispersion} is very close in both kinds of phase transitions, and agrees with the analytical estimate \eqref{eq:Dval} $D \approx 2.548$.

\section{Breakdown of quasistatic time evolution and scaling}\label{sec:breakdown}

In this section we depart from equilibrium and allow the temperature to change slowly in time. The background geometry is approximately \eqref{eq:adsBHmetric} with a time-dependent position of the horizon $z_h$, related to the temperature through \eqref{eq:temperature}. Applying the change of variables introduced at the beginning of section \ref{sec:fluct}, we arrived at \eqref{eq:scalareom}, where all explicit dependence on the horizon position has been removed from the equation of motion of the scalar. Thus, the time evolution is completely encoded in the boundary conditions of the scalar field. In the following we will restrict to homogeneous configurations and drop the terms with derivatives along the spatial field theory directions $x^i$.

\subsection{Quasistatic evolution}

If the change of the temperature is slow enough, the solution for the scalar field will in principle remain close to the static solution \eqref{eq:staticsol}, with small corrections proportional to the time derivative of the expectation value and/or temperature. The solution for the scalar field can be split in ``equilibrium'' and ``out-of-equilibrium'' terms,
\begin{equation}
\phi(v,Z)=\phi_0(v,Z)+\delta \phi(v,Z) \ ,
\end{equation}
where the equilibrium solution is taken to be
\begin{equation}
\phi_0(v,Z)=-\frac{\overline{\psi}(v)}{3}\eta_R(Z) \ ,\qquad \overline{\psi}=\left(\frac{4\pi T}{3} \right)^{-3/4}\psi \ .
\end{equation}
The out-of-equilibrium term can be expanded in terms proportional to time derivatives of the equilibrum solution
\begin{equation}
    \delta \phi(v,Z)\sim \partial_v \overline{\psi}\, \delta\phi^{(1)}(Z)+\partial_v^2 \overline{\psi} \,\delta\phi^{(2)}(Z)+\cdots,
\end{equation}
that can be solved iteratively oder by order in equation \eqref{eq:scalareom}.
Then, at the leading order in this expansion, the equation of motion for the out-of-equilibrium term is
\begin{equation}
\partial_Z \left( \frac{F(Z)}{Z^2}\partial_Z \delta\phi\right)+\frac{27/16}{Z^4}\delta\phi= \frac{1}{3}\partial_v\overline{\psi}(v)\left[ \frac{2}{Z^3}\eta_R(Z)-\frac{2}{Z^2} \eta_R'(Z)\right]+O(\partial_v^2 \overline{\psi})\ .
\end{equation}
The solution is of the same form as the one used to estimate the decay rate \eqref{eq:solchi1}, replacing $-\delta \overline{\psi}_0\Gamma\to \partial_v\overline{\psi}$. This allows for an immediate translation of the results in appendix \ref{app:decayrate} to this case. The leading and subleading terms in the boundary expansion of the solution are
\begin{equation}
\phi_-(v)=-\frac{1}{3}\overline{\psi} \ ,\qquad \phi_+(v)\approx \frac{\gamma}{3} \overline{\psi}+\frac{4}{9}\left( \frac{1}{2}-I_1\right)\partial_v \overline{\psi}+O(\partial_v^2 \overline{\psi}) \ .
\end{equation}
Expanded to lowest order in derivatives, the no source condition $J=0$ becomes
\begin{equation}
\frac{2}{9}\left(1-2I_1\right)\left(\frac{4\pi T}{3}\right)^{9/4}\partial_v\left[\left(\frac{4\pi T}{3}\right)^{-3/4}\psi\right]+ \partial_\psi V(\psi)\approx 0 \ .
\end{equation}
The expectation value is a small perturbation around the equilibrium value $\psi(v)=\psi_0[T(v)]+\delta \psi(v)$, satisfying $\partial_\psi V(\psi_0)=0$ and
\begin{equation}\label{eq:outofeqpert}
\delta\psi\approx \frac{\partial_v T}{\Gamma[\psi_0]}\left(\frac{3}{4 T}\psi_0-\partial_T\psi_0\right) \ ,
\end{equation}
where $\Gamma[\psi_0]$ is given in \eqref{eq:ratio}. Even if the temperature changes very slowly, $\partial_v T/T \ll 1$, when it approaches a critical or spinodal value, the assumption that $\delta \psi$ is small breaks down, since $\Gamma\to 0$. This shows that critical slowing down results in the breakdown of the quasistatic approximation.

\subsection{Critical and spinodal scaling}

Critical points of second-order phase transitions exhibit characteristic scaling properties that also affect the equilibrium states close to them. The correlation length and relaxation time diverge with particular exponents of the temperature difference close to the critical point, while the deviation from the critical values goes to zero:
\begin{equation}\label{eq::scalings}
 \xi\sim |T-T_c|^{-\nu} \ ,\qquad \tau\sim |T-T_c|^{-z\nu} \ ,\qquad \Delta \psi \sim |T-T_c|^\beta \ .
\end{equation}
Let us characterize the time evolution close to the critical point with a linear function
\begin{equation}
\frac{T_c-T}{T_c}=\frac{t}{\tau_{_Q}} \ ,
\end{equation}
where $\tau_{_Q}$ determines the rate of change of the temperature. The larger $\tau_{_Q}$ is, the slower the time evolution. We have selected $t=0$ as the point where the critical temperature is reached when one approaches it from from higher temperatures. At some point in the time evolution, the time remaining before reaching the critical point, $\Delta t = |t|$, will be the same as the relaxation time, $\Delta t=\tau(t)\equiv t_*$, so the system will not have time to equilibrate before undergoing the phase transition. This is known as the freeze-out time \cite{Zurek:1985qw}; note that by our definition, $t_*$ is positive even though the freeze-out occurs at negative times. We can use it, with the corresponding temperature $T_*=T(-t_*)$, to define when the quasistatic approximation breaks down and the system goes out of equilibrium. The relaxation time then satisfies
\begin{equation}
\tau=\tau_c\left(\frac{T_*-T_c}{T_c}\right)^{-z\nu}=\tau_c\left(\frac{t_*}{\tau_{_Q}}\right)^{-z\nu}=t_*  \ .
\end{equation}
From this, we obtain the scalings
\begin{equation}\label{eq:critscalingT}
t_*\sim \tau_{_Q}^{\frac{z\nu}{1+z\nu}},\ \ T_*\sim \tau_{_Q}^{-\frac{1}{1+z\nu}} \ .
\end{equation}
In the case of a closed system with a phase transition producing spontaneous symmetry breaking, this type of scaling determines the size of domains in the broken phase through the Kibble-Zurek mechanism \cite{Kibble1976,Zurek:1985qw}.

Our case differs in several ways from the type of transitions to which the Kibble-Zurek mechanism applies. In the first place, the two phases have the same symmetry realization, there is no spontaneous breaking of symmetry. Secondly, our initial state is homogeneous and fluctuations are completely suppressed, so there cannot be formation of domains. And lastly, we are describing an open system, since the sector undergoing the phase transition is coupled to a thermal bath.

Although the freeze-out argument is appealing, we need a more direct characterization of the departure from equilibrium. A simple choice is to compare the expectation value of the operator or the effective potential with their equilibrium values. This, however, leads to a different scaling exponent. Let us declare the quasistatic evolution to be valid as long as the deviation from the equilibrium value in \eqref{eq:outofeqpert} remains bounded and relatively small compared to the equilibrium value,
\begin{equation}\label{eq:outofeqcond}
\left|\frac{\delta \psi}{\psi_0}\right|<\epsilon<1 \ ,
\end{equation}
for some fixed $\epsilon$. 

Consider a first-order transition, and let $T_a$, $a=1,2$ be the two spinodal temperatures. We can expand close to the spinodal values, $\psi_0=\psi_a+\Delta\psi_a$, $T=T_a+\Delta T$. The following conditions are satisfied:
\begin{equation}
\partial_\psi V(\psi_0,T)=0=\partial_\psi V(\psi_a,T_a) \ ,\qquad \partial_\psi^2 V(\psi_a,T_a)=0 \ .
\end{equation}
From the condition that $\psi_0$ is an equilibrium value at temperature $T$ we obtain
\begin{equation}\label{eq:dpsia}
\Delta \psi_a\approx \pm (-\sigma_a \Delta T)^{1/2} \ ,\qquad \sigma_a=\frac{2\partial_T\partial_\psi V(\psi_a,T_a) }{\partial_\psi^3V(\psi_a,T_a)}  \ ,
\end{equation}
fixing the  value of the ``critical'' exponent $\beta=1/2$. Note that we can approach a given spinodal point only from higher or lower temperatures, but not from both. This is reflected in that $\Delta\psi$ is real only if $\Delta T$ has the correct sign. The dimensionless rate \eqref{eq:ratio} is approximately
\begin{equation}
\Gamma_a \approx \frac{9}{2(1-2I_1)} \left(\frac{4\pi T_a}{3}\right)^{-3/2} \partial_\psi^3V(\psi_a,T_a)\Delta \psi_a\ .
\end{equation}
In order to approach the point from a (meta)stable branch, we should select the sign such that $\partial_\psi^3 V(\psi_a,T_a) \Delta \psi_a>0$. Since $\Gamma_a$ is the inverse of the relaxation time, this gives  ``critical'' exponent $z\nu=1/2$ for the spinodal point.

Considering instead a second-order phase transition, close to the critical temperature $T=T_c+\Delta T$, $\psi_0=\psi_c+\Delta \psi_c$, the conditions on the derivatives of the effective potential are
\begin{equation}
 \partial_\psi V(\psi_0,T)=0=\partial_\psi V(\psi_c,T_c)\ ,\qquad \partial_\psi^2 V(\psi_c,T_c)=0=\partial_\psi^3V(\psi_c,T_c) \ .
\end{equation}
Now the temperature dependence appears with a modified exponent
\begin{equation}\label{eq:dpsic}
\Delta \psi_c\approx (-\sigma_c \Delta T)^{1/3} \ ,\qquad \sigma_c=\frac{6\partial_T\partial_\psi V(\psi_c,T_c) }{\partial_\psi^4V(\psi_c,T_c)}  \ ,
\end{equation}
fixing the critical exponent $\beta=1/3$. In this case we can always find a real root for $\Delta \psi_c$. The dimensionless  rate \eqref{eq:ratio} can be approximated by
\begin{equation}
\Gamma_c \approx \frac{9}{4(1-2I_1)}  \left(\frac{4\pi T_c}{3} \right)^{-3/2}\partial_\psi^4V(\psi_c,T_c)(\Delta \psi_c)^2 \ .
\end{equation}
Identifying $\Gamma_c$ with the inverse of the relaxation time results in the exponents $z\nu=2/3$ for the critical point of the second-order phase transition. Although the values for the exponents might look unusual, they are mean field values for the effective potential \eqref{eq:effpotT}. The results for the scaling exponents of both types of transitions in our model are summarized in Table~\ref{tab:critexp}.

\begin{table}[h!]
    \centering
    \begin{tabular}{|c||c|c|c|}
\hline        Order & \ $z$ \ & \ $\beta$ \ & \ $\nu$ \ \\
\hline       1st  & 2 & 1/2 & 1/4 \\
\hline  2nd & 2 & 1/3 & 1/3 \\ \hline
    \end{tabular}
    \caption{Scaling exponents as defined in \eqref{eq::scalings} for first and second-order phase transitions in the holographic model.}
    \label{tab:critexp}
\end{table}

From \eqref{eq:outofeqpert} and \eqref{eq:outofeqcond} we arrive at the following condition for the spinodal points:
\begin{equation}
\epsilon> \bar\tau_a\left|\frac{\partial_v T}{T-T_a} \right| \ .
\end{equation}
For a critical point we have instead
\begin{equation}
\epsilon> \bar \tau_c\left|\frac{ T_c^{1/3}\partial_v T}{(T-T_c)^{4/3}}\right| \ .
\end{equation}
In the above equations, we have defined
\begin{equation}
\bar\tau_a=\left(\frac{4\pi T_a}{3} \right)^{3/4}\left|\frac{1-2I_1}{9\partial_\psi^3V(\psi_a,T_a) }\right| \ ,\qquad \bar \tau_c=\left(\frac{4\pi T_c}{3} \right)^{-3/4}\left|\frac{4(1-2I_1)}{27(T_c\sigma_c)^{1/3} \partial_\psi^4V(\psi_c,T_c) }\right| \ .
\end{equation}
Then, assuming a linear dependence on time close to the spinodal/critical point $\Delta T_{a,c}=T_{a,c} t/\tau_{_Q}$, the time $t=-t_*$ when the system goes out of equilibrium before reaching the spinodal/critical point is
\begin{eqnarray}
\label{eq::scalingtimefirst}
(t_*)_a &\approx & \frac{3}{4\pi T_a}\frac{\bar \tau_a}{\epsilon}\sim \tau_{_Q}^0 \ ,\\ 
\label{eq::scalingtimesecond}
(t_*)_c &\approx& \left(\frac{3}{4\pi T_c}\right)^{3/4}\left(\frac{\bar\tau_c^3}{\epsilon^3} \tau_{_Q}\right)^{1/4}\sim \tau_{_Q}^{1/4} \ .
\end{eqnarray}
The temperature when this happens is
\begin{eqnarray}
\label{eq::scalingTempfirst}
\left|\frac{T_*-T_a}{T_a}\right|&\approx & \left( \frac{3}{4\pi T_a}\right)\frac{\bar\tau_a/\epsilon}{\tau_{_Q}}\sim \tau_{_Q}^{-1} \ ,\\
\label{eq::scalingTempsecond}
\left|\frac{T_*-T_c}{T_c}\right|&\approx & \left( \frac{3}{4\pi T_c}\right)^{3/4}\left(\frac{\bar\tau_c/\epsilon}{\tau_{_Q}} \right)^{3/4}\sim \tau_{_Q}^{-3/4} \ .
\end{eqnarray}
Although critical slowing down happens for both spinodal and critical points, the critical scaling found close to each type of point is different. In both cases it is different from \eqref{eq:critscalingT}. The general formulas for the scaling in terms of the critical exponents are
\begin{equation}
    t_*\sim \tau_{_Q}^{\frac{z\nu-\beta}{1+z\nu-\beta}} \ ,\qquad \left|\frac{T_*-T_c}{T_c}\right|\sim \tau_{_Q}^{-\frac{1}{1+z\nu-\beta}} \ ,\qquad (\beta<1) \ .
\end{equation}

\subsection{Full time evolution}
\label{sec:numericalresults}

So far we have used the properties of the effective potential and near-equilibrium solutions close to the critical/spinodal points  to estimate the approximate behaviour of the solutions as these points are slowly approached.  However, this does not give us precise quantitative information about how slow the evolution should be in order to enter the scaling regime before going out of equilibrium and how close to the spinodal/critical point this will happen.

In order to answer these questions we will now perform a full-fledged numerical calculation of the time evolution and compare the results against our analytical estimates of the scalings \eqref{eq::scalingtimefirst}-\eqref{eq::scalingTempsecond}.

In order to proceed we need to specify the time dependence of the temperature $T(v)$. We will pick as initial time of the evolution $v=0$, where the scalar solution will be equal to the equilibrium solution \eqref{eq:staticsol} at the initial temperature. In order for the equation of motion of the scalar to be satisfied at the initial time we also impose $\partial_v T(0)=0$. We also want a function that changes slowly to remain in the quasistatic regime and has a linear behaviour close to the time $v_{a,c}$ where the spinodal/critical temperature $T_{a,c}$ is reached, with a slope that can be tuned. 

A possible function satisfying these requirements is
\begin{equation}\label{eq:tempevol}
    \frac{T}{T_{\text{first}}} = \frac{T_{a,c}}{T_{\text{first}} } - \mathcal{B}\tanh\left[\frac{\sigma}{\mathcal{B}}(v-v_{a,c}) + \frac{\sigma}{2v_{a,c}\mathcal{B}}(v-v_{a,c})^2\right] \ ,
\end{equation}
with the parameter $\sigma$ determining the slope close to the spinodal/critical point and $\mathcal{B}$ determining the initial temperature
\begin{equation}
\tau_{_Q} =\frac{3}{4\pi  T_{\text{first}} \sigma}\,, \quad
    \frac{T(0)}{T_{\text{first}}}=\frac{T_{a,c}}{T_{\text{first}} }+\mathcal{B}\tanh\left(\frac{\sigma v_{a,c}}{2\mathcal{B}}\right)\ .
\end{equation}
In these formulas we use as a reference scale $T_{\text{first}} \approx 4.934M$, which is the critical temperature of the first-order transition corresponding to the first row in Table \ref{tab:parametVeff}. 

In order to avoid starting the evolution too close to the spinodal/critical point, we will require that as the evolution becomes slower $\sigma\ll 1$, the time to reach the spinodal/critical point is larger  $v_{a,c}\gg 1$. In addition, we will also require starting at a higher temperature so that, if the system wants to move away from equilibrium before reaching the spinodal/critical time, it has enough time to do so. This last condition requires $\sigma v_{a,c}/\mathcal{B} \gtrsim 1$ and $\mathcal{B}\gg 1$. These conditions can be achieved by imposing the following relations among the parameters
\begin{equation}
    \sigma = \frac{2}{v_{a,c}}\left(4\log_{10}v_{a,c}-\frac{T_{a,c}}{T_{\text{first}} }\right) \ , \qquad
    \frac{T(0)}{T_{\text{first}}}=2\log_{10}v_{a,c} \ .
\end{equation}
This leaves $v_{a,c}$ as the only parameter of the time evolution. 

The strategy we followed is, given a theory determined by the parameters in the effective potential, choose a set of increasing values for $v_{a,c}$, which corresponds to another set of increasing values of $\tau_{_Q}$. For each value, we numerically solve the equations as explained in appendix \ref{app::numericalcomputation} and define the time $t_*$ at which the system falls out-of-equilibrium by \eqref{eq:outofeqcond} for several sufficiently small values of $\epsilon$ that let us explore the scaling regime.

\subsubsection{Scaling in first and second-order phase transitions}

\begin{figure}[t!]
\begin{center}
{\includegraphics[width=0.49\textwidth]{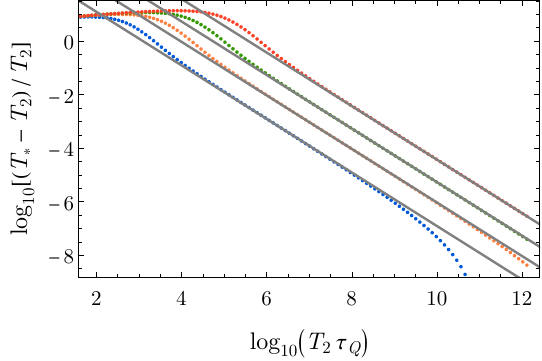}
\includegraphics[width=0.48\textwidth]{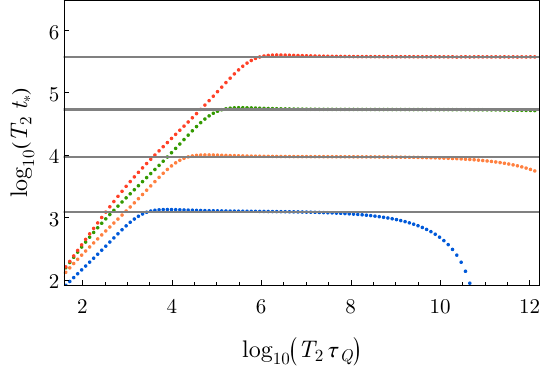}}
\end{center}
\caption{Scaling of $T_*$ and $t_*$ with the characteristic cooling rate $\tau_{_Q}$ for the first-order phase transition with parameters displayed in the first row of Table \ref{tab:parametVeff}. The different curves were obtained for values of epsilon equal to  $3\cdot10^{-4}$ (blue),
$4\cdot 10^{-5}$ (orange), $7\cdot 10^{-6}$ (green), and  $10^{-6}$ (red). The gray lines correspond to the theoretical expectation for the scaling of both time and temperature for a first-order phase transition.}
\label{fig::scalingsfirstordercaseNew}
\end{figure}

\begin{figure}[t!]
\begin{center}
{\includegraphics[width=0.49\textwidth]{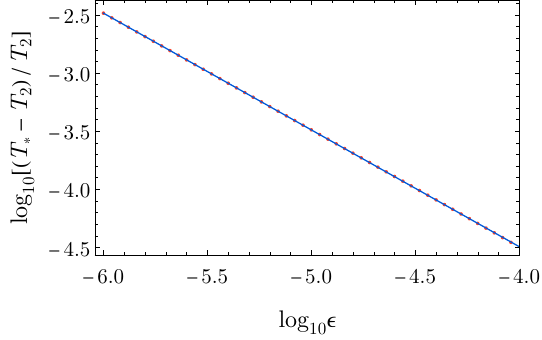}
\includegraphics[width=0.48\textwidth]{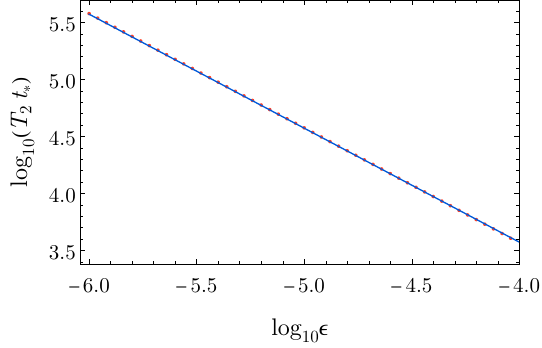}}
\end{center}
\caption{First-order phase transition (first row of Table \ref{tab:parametVeff}). Scaling of $T_*$ and $t_*$ with $\epsilon$ for time evolution 
for $\tau_Q \approx 1.16 \cdot 10^8 T_2^{-1}$.
The red dots correspond to the data extracted from the numerical evolutions. The blue lines correspond to the theoretical curves.}
\label{fig::scalingsdeltafirstordercase}
\end{figure}

For the case of the first-order phase transition with parameters as in the first row of Table \ref{tab:parametVeff}, we obtained the results shown in Figure  \ref{fig::scalingsfirstordercaseNew} for a few choices of $\epsilon$. In gray we added the line corresponding to the scaling that we estimated analytically for a first-order phase transition \eqref{eq::scalingtimefirst}, \eqref{eq::scalingTempfirst}. We observe that all curves show a scaling compatible with the analytical estimate, being its associated range in $\tau_{_Q}$ larger for smaller values of $\epsilon$. As $\epsilon$ decreases, the scaling regime will start at larger values of $\tau_{_Q}$ while deviations from it will take place at even larger values. Additionally, the time to fall out-of-equilibrium $t_*$ increases, letting the system probe the regime closer to the spinodal point. 
Furthermore, in Figure~\ref{fig::scalingsdeltafirstordercase}, we show that the scaling of the temperature and the time is compatible with the prediction at fixed $\tau_{_Q}$ and as a function of $\epsilon$. The deviations observed at large values of $\tau_{_Q}$ are related to the finite resolution of the spinodal point determined by $\epsilon$. Namely, we cannot observe deviations from equilibrium if the evolution is so slow that they happen at separations from the spinoidal point smaller than $\epsilon$.

This is consistent with the scaling region becoming larger at smaller $\epsilon$. As a consequence of this limitation in resolution, for larger values of $\epsilon \gtrsim 10^{-3}$ the scaling region ceases to be observable. 
Taking this into account, we see that the cooling rate should be rather small in order for the scaling to be observed $1/\tau_{_Q}\lesssim (10^{-6} - 10^{-8}) T_2$, depending on $\epsilon$. On the other hand, the temperature where the scaling regime is entered does not seem to be as sensitive to the value of $\epsilon$ as the cooling rate. Indeed, declaring that the scaling region starts when the deviation of the temperature scaling from the theoretical value is $10 \%$, for all the cases studied in Figure~\ref{fig::scalingsfirstordercaseNew}, we  find that the scaling regime starts when $(T_*-T_2)/T_2 \approx 0.07 - 0.08$.

\begin{figure}[t!]
\begin{center}
{\includegraphics[width=0.49\textwidth]{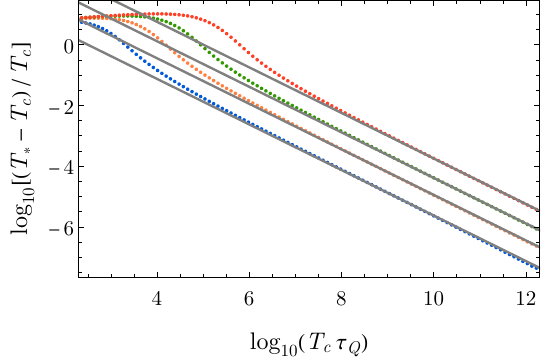}
\includegraphics[width=0.48\textwidth]{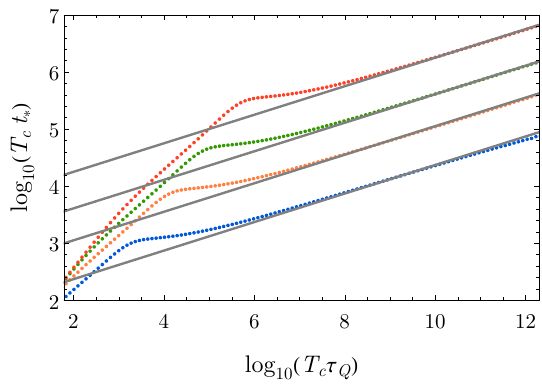}}
\end{center}
\caption{Scaling of $T_*$ and $t_*$ with the characteristic cooling rate $\tau_{_Q}$ for the second-order phase transition with parameters displayed in the second row of Table \ref{tab:parametVeff}. The different curves were obtained for values of epsilon equal to  $3\cdot10^{-4}$ (blue),
$4\cdot 10^{-5}$ (orange), $7\cdot 10^{-6}$ (green), and  $10^{-6}$ (red). The gray lines correspond to the theoretical expectation for the scaling of both time and temperature for a second-order phase transition.}
\label{fig::scalingssecondordercaseNew}
\end{figure}

\begin{figure}[t!]
\begin{center}
{\includegraphics[width=0.49\textwidth]{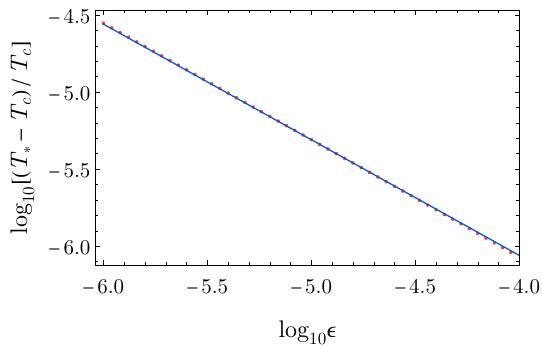}
\includegraphics[width=0.48\textwidth]{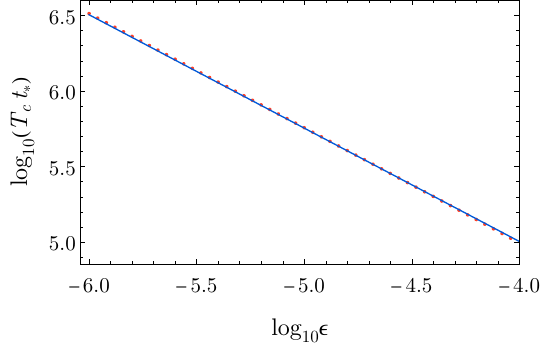}}
\end{center}
\caption{Second-order phase transition (second row of Table \ref{tab:parametVeff}). Scaling of $T_*$ and $t_*$ with $\epsilon$ for time evolution with $\tau_Q=1.16\cdot 10^{11} T_c ^{-1}$.
The red dots correspond to the data extracted from the numerical evolutions. The blue lines correspond to the theoretical curves.}
\label{fig::scalingsdeltasecondordercase}
\end{figure}

In the case of the second-order phase transition, whose parameters are given in the second row of Table \ref{tab:parametVeff}, we expect to find a different scaling, given in \eqref{eq::scalingtimesecond}, \eqref{eq::scalingTempsecond}. We show the results for the same choices of $\epsilon$ as earlier in Figure \ref{fig::scalingssecondordercaseNew}, in which the gray lines represent the scaling estimated analytically. We observe that our numerical results exhibit scaling regimes and agree with the analytical estimate. Once again, as $\epsilon$ is decreased, the system requires greater $\tau_{_Q}$ to enter the scaling region. However, this time there are no clear signs of deviations for the three smaller choices of $\epsilon$ for the largest $\tau_{_Q}$ shown here. 
Similarly to the previous case, we find a compatible scaling with $\epsilon$ at fixed $\tau_{_Q}$, as shown in Figure \ref{fig::scalingsdeltasecondordercase}.

Compared to the first-order phase transition, the necessary cooling rate to observe the scaling region is still comparably small $1/\tau_Q \lesssim (10^{-7}-10^{-9}) T_c$, although the scaling region is still relatively large for the larger value of $\epsilon$ shown in Figure~\ref{fig::scalingssecondordercaseNew}. The temperature seems also to have a weak dependence on $\epsilon$ with the onset of the scaling regime closer to the critical temperature than in the first-order phase transition. Using the same criterion as for the first-order phase transition, we find that for all the cases shown in Figure \ref{fig::scalingssecondordercaseNew}
the scaling region starts at $(T_*-T_c)/T_c \approx 0.01$.

\subsubsection{Scaling in crossovers and weak first-order phase transitions}

\begin{figure}[t!]
\begin{center}
{\includegraphics[width=0.48\textwidth]{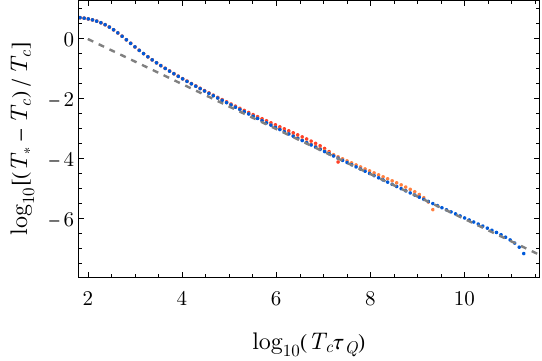}
\includegraphics[width=0.48\textwidth]{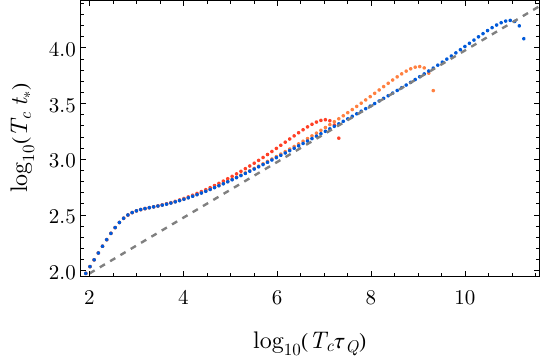}}
\end{center}
\caption{Scaling of $T_*$ and $t_*$ with the characteristic cooling rate $\tau_{_Q}$ for the crossovers with $A= 5 \cdot 10^{-3}$ (red),
$A= 5 \cdot 10^{-4}$ (orange), and $A= 5 \cdot 10^{-5}$ (blue).
Here, $\epsilon = 10^{-3}$. The gray line corresponds to the theoretical expectation for the scaling of both time and temperature for a second-order phase transition.}
\label{fig::scalingscrossoverNew}
\end{figure}

As the second-order phase transition is transformed into a crossover, we expect to stop observing a scaling regime. However, as long as the crossover is strong enough, one could in principle observe some approximate scaling in a limited region of cooling rates. This is displayed in Figure \ref{fig::scalingscrossoverNew}, where we show time and temperature in which the system falls out-of-equilibrium as a function $\tau_{_Q}$ for two different crossovers, given by the parameters in the third and fourth rows of Table~\ref{tab:parametVeff},
and for a single choice of $\epsilon = 10^{-3}$.

We observe that the stronger crossover still exhibits a region at large values of $\tau_{_Q}$ in which there is an approximate scaling not far from the one expected for a second-order phase transition. The behaviour of $T_*$ and $t_*$ for the weaker crossover closely follows the stronger one, but noticeable departures from it take place at around $\tau_{_Q} \sim 10^7 T_c^{-1}$, not exhibiting a clear scaling regime.

As $\tau_{_Q}$ increases, a big drop in the values of $t_*$ and $T_*$ happens and deviation from equilibrium are not observed any further. The reason behind it is that, as the second derivative of the potential does not vanish at $T_c$, we can always remain closer and closer to equilibrium in an evolution by increasing more and more $\tau_{_Q}$. With the specified criterion of $\epsilon = 10^{-3}$, the system remains always in equilibrium for larger values of $\tau_{_Q}$ than the ones for which we displayed data in Figure \ref{fig::scalingscrossoverNew}. We checked that decreasing $\epsilon$ increases the range of cooling rate for which we find data, but this range always remains finite. As it is expected, we find data for a larger range of $\tau_{_Q}$ for the stronger crossover as it is closer to a second-order phase transition.

A related interesting case arises when considering weak first-order phase transitions, meaning those that are close to being a second-order one. In such case one would expect to find a first-order phase transition scaling regime for slow enough evolution of the temperatures. However, at intermediate cooling rates, when the system falls out-of-equilibrium not too close to the spinodal point, one may expect to find a scaling close to that of a second-order phase transition if the transition is weak enough. The three examples we will consider to explore this possibility are defined by the parameters given in the last three rows of Table~\ref{tab:parametVeff}.

\begin{figure}[t!]
\begin{center}
{\includegraphics[width=0.46\textwidth]{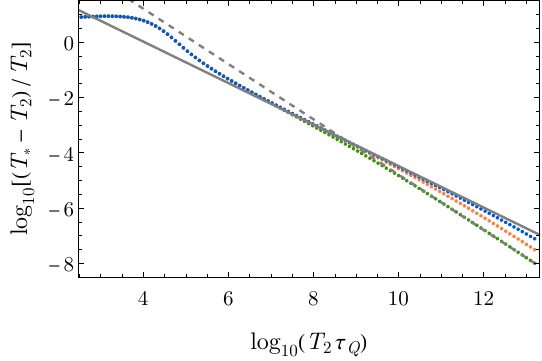}
\includegraphics[width=0.46\textwidth]{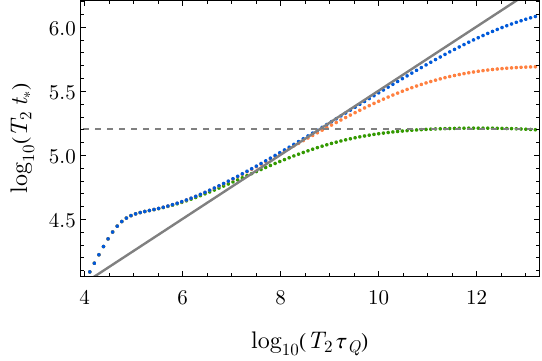}}
\end{center}
\caption{Scaling of $T_*$ and $t_*$ with the characteristic cooling rate $\tau_{_Q}$ for the weak phase transition with $A=5 \cdot 10^{-3}$ (green), $A=5 \cdot 10^{-4}$ (orange), and $A=5 \cdot 10^{-5}$ (blue). For all the curves, $\epsilon = 10^{-5}$. The gray line correspond to the theoretical expectation for the scalings of a first-order (dashed) and second-order phase (solid) transitions.}
\label{fig::scalingsFOPTWeakNew}
\end{figure}

Figure \ref{fig::scalingsFOPTWeakNew} shows the dependence of the time and temperature in which the field falls out of equilibrium for the choice of $\epsilon = 10^{-5}$ and for the three mentioned theories. We have added for visual aid a solid and dashed gray lines representing the second and first-order scalings respectively. These lines were forced to pass through the curve for the weakest and strongest transitions respectively.

The results show that for fast coolings, all three curves agree and seem to start following the second-order scaling (solid gray line). The weaker the transition, the closer it gets to exhibit such scaling, showing deviations for larger values of $\tau_{_Q}$. These deviations at large $\tau_{_Q}$ seem to be getting the curves towards the first-order scaling (dashed gray line). For the strongest transition we studied (green dots), the first-order scaling does clearly exhibit itself in the studied range of $\tau_{_Q}$. The mid strength transition results (orange dots) show the scaling at the very last values of cooling rate, being this a region that is likely prolonged toward higher values of $\tau_{_Q}$.

Our results are compatible then with the idea that weak first-order phase transition exhibit a second-order scaling before showing the first-order one for very slow temperature evolution.

\begin{figure}[t!]
\begin{center}
{\includegraphics[width=0.46\textwidth]{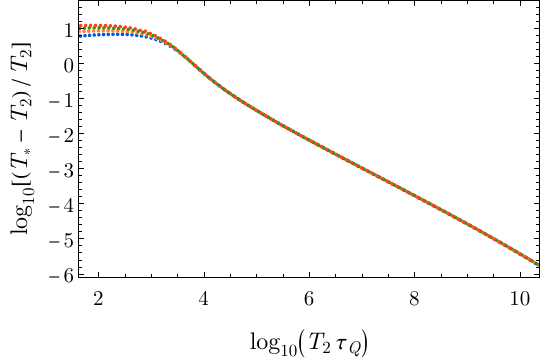}
\includegraphics[width=0.46\textwidth]{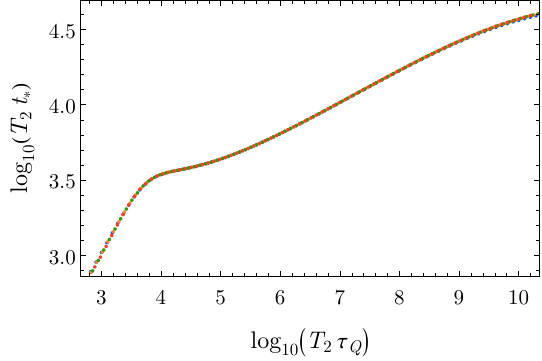}}
\end{center}
\caption{Comparison of $T_*$ and $t_*$ as a function of the cooling rate $\t_Q$ for the weak first-order phase transition with $A = 5 \cdot 10^{-4}$  for different choices of epsilon: 
$10^{-4}$ (blue), $10^{-5}$ (orange), $10^{-6}$ (green), and $10^{-7}$ (red). We have performed a shift in both axes for the curves to overlap.}
\label{fig::comparisondeltaFOPTMid}
\end{figure}

Finally, Figure \ref{fig::comparisondeltaFOPTMid} compares $T_*$ and $t_*$ for different values of $\epsilon$ for the first-order phase transition with $A=5 \cdot 10^{-4}$ (see Table \ref{tab:parametVeff}). In order for the curves to overlap, we have made a shift in both axes so that the intermediate region falls in the same range. 

We observe that all curves lie on top of each other. This implies that the intermediate region of the curves, before the first-order phase scaling arises, follows an equivalent behaviour and changing $\epsilon$ does not help finding closer scalings to a second-order phase transition. The only change that the choice of $\epsilon$ produces is the displacement of the region in which the first-order scaling emerges to larger values of $\tau_{_Q}$.

In conclusion, our results show that the proximity to a second-order phase transition is the main factor determining whether an intermediate second-order scaling region manifests itself.

\section{Discussion and outlook}\label{sec:discussion}

Our results highlight similarities in the behaviour of a system when driven slowly through a first- or a second-order phase transition, if fluctuations are suppressed: In both cases there is critical slowing down of fluctuations and the departure from equilibrium obeys a scaling law related to the equilibrium ``critical'' exponents of critical and spinodal points. In addition, we have obtained specific approximate values for the relative temperature when the scaling regimes are entered, and it will be interesting to explore how much these are model dependent or whether they hold more generally than the cases we have studied.  

In practice, whether the scaling regime can be reached will depend on the relative suppression of fluctuations, which depend on the properties of the particular system undergoing the transition. However, the fact that spinodal scaling has been observed in experiments shows that understanding this regime is not a purely academic exercise, but that can also be relevant for real systems.

It should be noted that there can be different behaviours to the one presented here for a first-order phase transition when it is the volume of a closed system rather than the temperature of an open system the parameter that is evolving. In the first case the energy density can be equivalently used as time-evolving parameter so that, in contrast to the open system, there is a single homogeneous equilibrium state at each time and in principle the unstable state lying between the spinodal points could be reached.

There are several interesting directions one can pursue using the simple holographic model described here:
\begin{itemize}
\item Although we have studied only the case of undercooling, one can similarly study the complementary overheating case, although we expect no qualitative differences.
    \item We have focused on the approach towards the spinodal/critical points. Evolving further in time, one could study possible freeze-out behavior, as well as the approach to the new equilibrium state.
    \item Our model could easily be adapted to phase transitions involving spontaneous symmetry breaking. A crucial difference with the transitions we have studied here is that the symmetric high temperature state is still present as an unstable state at low temperatures. The non-trivial evolution to the ground state then happens through (small, random) perturbations. Adding spatial dependence, one could realize the Kibble-Zurek mechanism of domain formation.
    \item Adding spatial dependence to our field, there would be other interesting inhomogeneous configurations whose time-evolution could be studied, including the expansion of bubbles \cite{Bea:2021zsu} and spinodal decomposition \cite{Bea:2021zol}.
    \item One could simulate fast, instead of slow, quenches\footnote{Quenches that are fast compared to the relaxation rate of the scalar field but that satisfy $T'(t)/T^2\ll 1$ can be captured with the formalism presented here. Quenches not far from the spinodal point are good candidates to work in this regime as $\Gamma_T \ll T$. Alternatively, one can realize fast quenches by fixing the temperature and making the couplings of the theory time-dependent.}. In particular, one could ask if the system can transition {\it before} reaching the spinodal point, if the evolution is fast enough to give it sufficient energy to jump the potential barrier. That is, could the phase transition be completed while the false vacuum is still metastable, not by bubble nucleation but using energy pumped in during the quench? 
    \item The model could also be used to study dynamical phase transitions  \cite{Heyl:2017blm} and Floquet states \cite{Tsuji:2023dar} \footnote{See \cite{Biasi:2017kkn,Ishii:2018ucz,Garbayo:2020dmh,Berenguer:2022act} for holographic realizations of Floquet states.} by turning on an oscillating source. It could be particularly interesting when the oscillation happens around the critical or spinodal points. 
    \item In previous work, some of us have discussed how to use holography to compute the field theory's effective action in a derivative expansion \cite{Ares:2021ntv}. This could easily be done in the current model; it would then be interesting to study how well the numerical evolution reported here could be captured by solving the equation of motion coming from the effective action (which is an ordinary differential equation instead of a partial one).
\end{itemize}

\newpage
\begin{itemize}
    \item Instead of a black hole background geometry one could consider an AdS soliton geometry or a global AdS geometry. In this case the time evolution of the confinement scale could produce quantum phase transitions. 
\end{itemize}

In addition to the points listed above, there are several possible extensions and generalizations of the holographic model that could be interesting to consider and for which similar questions can also be addressed:

Phase transitions involving a single scalar could also be realized by introducing a non-trivial potential in the gravitational action, rather than by modifying boundary conditions as done in this work, see e.g. \cite{Bea:2018whf}. 

Treating the background evolution non-adiabatically would allow to consider more general evolutions than the ones mentioned earlier, with which to probe the far-from-equilibrium regime. Widely used examples of time-evolving analytic backgrounds are the dual to a boost invariant plasma \cite{Janik:2005zt,Heller:2008mb} and Vaidya geometries used to describe thermalization \cite{Bhattacharyya:2009uu}. 

Finally, going beyond the probe approximation and taking into account the backreaction of the scalar over the metric will in general modify the effective potential and even could result in the introduction of non-analytic terms, allowing for a generalization to higher dimensions as in e.g.\cite{Ares:2021ntv}. Adding other types of fields, such as charged scalars and gauge fields will extend the phase diagram and allow for transitions produced by changes in the chemical potential rather than the temperature.

\section*{Acknowledgments}
We would like to thank Jorge Casalderrey Solana, Mark Hindmarsh, Javier Mas,  David Mateos and Alfonso Ramallo for valuable discussions and useful comments. The work of O.H. is supported by the Research Council of Finland (grant number 330346) and the Waldemar von Frenckell foundation. The work of A.C. is supported by Ministerio de Ciencia e Innovación de España under the program Juan de la Cierva-formación. A.C. and C.H. are partially supported by the AEI and the MCIU through the Spanish grant
PID2021-123021NB-I00.
The work of M.S.G. is supported by the European Research Council (ERC) under the European Union's Horizon 2020 research and innovation program (grant agreement No758759).

\appendix

\section{Stationary points of the effective potential}\label{app:criticalpoints}

Here we study in greater detail the equilibrium states of our system; these can be extracted from the effective potential \eqref{eq:effpotT}, which we reproduce here for convenience:
\begin{equation}
     V(\psi)=\Lambda \psi+\frac{1}{2}a_T\psi^2+\frac{b}{3}\psi^3+\frac{c}{4}\psi^4  \qquad \text{with} \qquad a_T=a+\frac{\gamma}{3}\left(\frac{4\pi T}{3}\right)^{3/2} \ .
\end{equation}
Firstly, we take all coefficients $\{\Lambda,a,b,c\}$ real, and $c>0$ in order to avoid a runaway behaviour of the potential. The no-source boundary condition \eqref{eq:criticalcond} has at least one and at most three solutions for $\psi$ real. If there is only one solution, the effective potential has one minimum, while if there are three real solutions the potential has two minima and a maximum. Two solutions correspond to limiting cases where the maximum coincides with a minimum and the potential has an inflection point. 

It will be convenient to write all dimensionful quantities in units of a reference mass scale $M$:
\begin{equation}
    a_T = \bar{a}_T M^{3/2} \ , \qquad  b = \bar{b} M^{3/4} \ ,  \qquad  \Lambda = \bar{\Lambda} M^{9/4} \ , \qquad \psi=\bar{\psi}M^{3/4} \ .
\end{equation}
The cubic equation \eqref{eq:criticalcond} has three real roots when the discriminant is positive. In terms of the dimensionless parameters the discriminant becomes
\begin{equation}
    \text{Discr} = \frac{1}{27 c^3}\left[4(\bar{b}^2  - 3 \bar a_T c)^3 - (2 \bar{b}^3 - 9 \bar{a}_T \bar{b} c + 27 c^2 \bar{\Lambda} )^2\right] \ .
\end{equation}
We see that for $T/M\to \infty$,  $\bar{a}_T\sim (T/M)^{3/2}\to \infty$ and $\text{Discr}\sim -\bar{a}_T^3$, so at high temperatures the discriminant is always negative, meaning that there is a single equilibrium state. When we lower the temperature there is a possibility that the discriminant becomes positive. This requires
\begin{equation}\label{eq:discrcond}
    \bar{b}^2  - 3 \bar a_T c>0 \ ,\qquad  (\bar{b}^2  - 3 \bar a_T c)^3\geq \frac{1}{4}(2 \bar{b}^3 - 9 \bar{a}_T \bar{b} c + 27 c^2 \bar{\Lambda} )^2 \ .
\end{equation}
Let us identify $T_1$ as the temperature where a second real root first appears. We introduce the parametrization
\begin{equation}
\label{eq:parameterizationofpotentialparameters}
    a=-\frac{\gamma}{3}\left(\frac{4\pi T_1}{3}\right)^{3/2}+\frac{b^2}{3 c}(1-A) \ ,\qquad  \Lambda=\frac{b^3}{27 c^2}\pq{1-A (B+2)} \ .
\end{equation}
Then, at $T=T_1$,
\begin{equation}
    \bar{a}_{T_1}=\frac{\bar{b}^2}{3 c}(1-A) \ ,
\end{equation}
which gives the condition
\begin{equation}
    A^2\pq{4 A-(B-1)^2}=0 \ .
\end{equation}
So we should impose $A=(B-1)^2/4$. Let us now identify $T=T_2<T_1$ as the temperature where two of the real roots coincide before moving to the complex plane and leaving a single real root. In this case
\begin{equation}
    \bar{a}_{T_2}=\frac{\bar{b}^2}{3 c}(1-A-\Delta A) \ ,
\end{equation}
where we have introduced a parametrization
\begin{equation}
    \frac{\gamma}{3}\left(\frac{4\pi}{3}\right)^{3/2}\frac{T_1^{3/2}-T_2^{3/2}}{M^{3/2}}=\frac{\bar{b}^2}{3 c} \Delta A>0 \ .
\end{equation}
The second condition in \eqref{eq:discrcond} is saturated when $\Delta A=0$, and for
\begin{equation}
    \Delta A_\pm=\frac{3}{4}+\frac{3}{8}B(2-B)\pm \frac{B+2}{8}\left[3(4-B^2) \right]^{1/2} \ .
\end{equation}
The range where there are two positive solutions $\Delta A_\pm>0$ is $1<B<2$. The temperature $T_2$ corresponds to $\Delta A_-$, and we must demand that
\begin{equation}
   \frac{\bar{b}^2}{3 c} \Delta A_+> \frac{\gamma}{3}\left(\frac{4\pi}{3}\right)^{3/2}\frac{T_1^{3/2}}{M^{3/2}}, \quad \Rightarrow \quad  \frac{\bar{b}^2}{3 c} (\Delta A_+-\Delta A_-)> \frac{\gamma}{3}\left(\frac{4\pi}{3}\right)^{3/2}\frac{T_2^{3/2}}{M^{3/2}} \ ,
\end{equation}
so that there is a single real root in the interval $T_2>T\geq 0$.

In the case where there is a second-order phase transition the potential develops an inflection point but there is always a single minimum. This corresponds to the discriminant having a double root, which happens at the extreme of the allowed interval for $B$: $B=1$, $A=\Delta A_-=0$. The critical temperature can be identified as $T_c=T_1$, and should satisfy the constraint
\begin{equation}\label{eq:secondordTccons}
    \frac{9}{4}\frac{\bar{b}^2}{3 c} > \frac{\gamma}{3}\left(\frac{4\pi}{3}\right)^{3/2}\frac{T_c^{3/2}}{M^{3/2}} \ .
\end{equation}
A crossover can be obtained by taking $B=1$ and $A<0$ but small, such that the discriminant remains negative at the would-be double root. The smaller $A$, the more pronounced the crossover would be. In this case $T_{\rm cross}=T_1$ would serve as a crossover temperature, with a constraint similar to \eqref{eq:secondordTccons} save for a small correction proportional to $A$.

\section{Analytic estimate of the decay rate and correlation length}\label{app:decayrate}

In this appendix we estimate the decay rate in amplitude of the perturbation and the correlation length. Let us introduce the differential operator acting on static solutions
\begin{equation}
{\cal D}_Z\,\phi\equiv \partial_Z \left( \frac{F(Z)}{Z^2}\partial_Z \,\phi\right)+\frac{27/16}{Z^4}\phi \ .
\end{equation}
The associated Green's function is
\begin{equation}
{\cal D}_ZG(Z,Z')=\delta(Z-Z'),\ \ G(Z,Z')=-\frac{2}{3}\left\{ \begin{array}{lc} \eta_R(Z)\eta_{9/4}(Z'), &Z'<Z \\ \eta_{9/4}(Z)\eta_R(Z'), & Z<Z'\end{array}\right.
\end{equation}
Where the solutions $\eta_{9/4}(Z), \eta_R(Z)$ were defined in \eqref{eq:scalarsols} and \eqref{eq:regularsol}.

\subsection{Decay rate}

Using the Green's function the equation for a homogeneous perturbation can be recast as the integral equation
\begin{equation}
\delta \phi=-\frac{1}{3}\delta\overline{\psi}(v)\eta_R(Z)-2\int_{z_{_{UV}}}^1 dZ' \, G(Z,Z') \left(\frac{1}{{Z'}^3}\partial_v\delta\phi-\frac{1}{{Z'}^2} \partial_{Z'}\partial_v \delta \phi \right) \ .
\end{equation}
We have introduced a cutoff $0<z_{_{UV}}\ll 1$ in the integration for convenience; we will take $z_{_{UV}}\to 0$ at the end of the calculation.

We can assume an exponential decay of the perturbation
\begin{equation}
\delta \phi(Z,v)=e^{-\Gamma v} \delta \phi_0(Z) \ ,\qquad \delta\overline{\psi}(v)=e^{-\Gamma v}\delta\overline\psi_0 \ .
\end{equation}
In this case, the integral equation simplifies to
\begin{equation}
\delta \phi_0(Z)=-\frac{1}{3}\delta\overline{\psi}_0\eta_R(Z)+2\Gamma\int_{z_{_{UV}}}^1 dZ' \, G(Z,Z') \left(\frac{1}{{Z'}^3}\delta\phi_0 -\frac{1}{{Z'}^2} \partial_{Z'}\delta \phi_0\right) \ .
\end{equation}
In principle a solution could be found using Neumann's series
\begin{equation}
\delta \phi_0(Z)=\sum_{n=0}^\infty \Gamma^n \chi_n(Z) \ .
\end{equation}
The first two orders in the expansion give 
\begin{equation}
\chi_0(Z)=-\frac{1}{3}\delta\overline{\psi}_0\eta_R(Z) \ ,
\end{equation}
and
\begin{equation}\label{eq:solchi1}
\chi_1(Z)=-\frac{2}{3}\delta\overline{\psi}_0\Gamma\int_{z_{_{UV}}}^1 dZ' \, G(Z,Z') \left[\frac{1}{{Z'}^3}\eta_R(Z') -\frac{1}{{Z'}^2} \eta_R'(Z')\right] \ .
\end{equation}
We will need the $z_{_{UV}}\to 0$ expansion of the integral,
\begin{equation}
\int_{z_{_{UV}}}^1 dZ'  \eta_R(Z')\left[\frac{1}{{Z'}^3}\eta_R(Z')-\frac{1}{{Z'}^2} \eta_R'(Z') \right]\simeq -\frac{1}{2}(1-z_{_{UV}}^{-1/2})+I_1+O(z_{_{UV}}) \ ,
\end{equation}
where $I_1 \approx 0.141115$.

We now expand the solution for $Z=z_{_{UV}}\to  0$ and identify the coefficients of the leading and subleading solutions
\begin{equation}
\begin{split}
\chi_0(z_{_{UV}}) &\simeq -\frac{1}{3}\delta\overline{\psi}_0 \left(z_{_{UV}}^{3/4}-\gamma z_{_{UV}}^{9/4}\right)+\cdots \ ,\\
\chi_1(z_{_{UV}}) &\simeq \frac{4}{9}\delta\overline{\psi}_0\Gamma\left[\frac{z_{_{UV}}^{7/4}}{2}-  \left(\frac{1}{2}-I_1\right)z_{_{UV}}^{9/4}\right]+\cdots \ .
\end{split}
\end{equation}
From this we can read the coefficient of the subleading solution
\begin{equation}
\delta \overline{\phi}_+(v)\simeq \left[] \frac{\gamma}{3}-\frac{4\Gamma}{9}\left(\frac{1}{2}-I_1\right)\right]\delta\overline{\psi}(v) \ .
\end{equation}
Then, imposing $\delta J=0$ in \eqref{eq:deltaJ}, we arrive at
\begin{equation}\label{eq:bcpert}
\left[] \frac{\gamma}{3}-\frac{4\Gamma}{9}\left(\frac{1}{2}-I_1\right)\right]\left(\frac{4\pi T}{3} \right)^{3/2}+W''(\psi_0)=0 \ .
\end{equation}
Let us note now that, at equilibrium,
\begin{equation}
0=J_0=V'(\psi_0)=\Lambda+W'(\psi_0)+\frac{\gamma}{3}\left(\frac{4\pi T}{3} \right)^{3/2}\psi_0 \ .
\end{equation}
Then, we can arrange terms in \eqref{eq:bcpert} such that the condition boils down to
\begin{equation}
-\frac{4\Gamma}{9}\left(\frac{1}{2}-I_1\right)\left(\frac{4\pi T}{3} \right)^{3/2}+V''(\psi_0)=0 \ .
\end{equation}
Hence, the decay rate is determined by
\begin{equation}\label{eq:ratioapp}
\Gamma=\frac{9}{2(1-2I_1)}\left(\frac{4\pi T}{3} \right)^{-3/2}V''(\psi_0) \ .
\end{equation}

\subsection{Correlation length}

We can restrict to time-independent solutions. For small spatial momentum the zero frequency plane wave perturbation \eqref{eq:planewave} can be expanded as
\begin{equation}
\delta\phi_k(Z)=\delta\phi_k^{(0)}(Z)+\bar k^2 \delta \phi_k^{(1)}(Z)+O(k^4) \ .
\end{equation}
The equations for the first two terms in the expansion are
\begin{eqnarray}
    {\cal D}_Z \delta \phi_k^{(0)} &=& 0 \ ,\\
    {\cal D}_Z \delta \phi_k^{(1)} &=& \frac{1}{Z^2}\delta \phi_k^{(0)} \ ,
\end{eqnarray}
with the solutions
\begin{eqnarray}
    \delta \phi_k^{(0)}(Z) &=& -\frac{\delta \bar\psi_k}{3}\eta_R(Z) \ ,\\
    \delta \phi_k^{(1)}(Z) &=& \int_0^1 \frac{dZ'}{{Z'}^2} G(Z,Z')\delta \phi_k^{(0)}(Z') \ .
\end{eqnarray}
In this case a radial cutoff is not necessary since the integral with the Green's function remains finite. The near-boundary expansion of the solution to this order is
\begin{equation}
    \delta \phi_k(Z)\sim -\frac{\delta\bar\psi_k}{3}\left[Z^{3/4}-\left( \gamma+\frac{2I_2}{3}\right)Z^{9/4} \right] \ ,
\end{equation}
where
\begin{equation}
    I_2=\int_0^1 dZ' \left(\frac{\eta_R(Z')}{Z'}\right)^2\approx 1.829 \ .
\end{equation}
Reading the asymptotic coefficients from the expression above, the no source condition \eqref{eq:deltaJ} becomes
\begin{equation}
    \left(V''(\psi_0)+\frac{2I_2}{9}\left( \frac{3}{4\pi T}\right)^{1/2}k^2\right)\delta \psi_k=0 \ .
\end{equation}
This is only satisfied for a purely imaginary value of the momentum that determines the correlation length
\begin{equation}
    -k^2=\xi^{-2}=\frac{9}{2I_2}\left( \frac{4\pi T}{3}\right)^{1/2}V''(\psi_0) \ .
\end{equation}

\section{Quasinormal modes}\label{app:quasinormalmodes}
In this appendix, we provide some details on the computation of the quasinormal modes of our model. These modes describe 
the decay of the scalar field fluctuations close to the spinodal points studied in section \ref{subsec::criticalslowingdown}.

We did this for both homogeneous, and non-homogeneous fluctuations satisfying the scalar equation \eqref{eq:pertscalareom} and use the plane wave expansion \eqref{eq:planewave}. 
We impose ingoing conditions at the horizon $Z=1$ and the no-source boundary condition $\delta J=0$ at $Z=0$. In the following, it will be convenient to adopt the coordinate $u=Z^{1/2}$ and to introduce the rescaled scalar field $\Phi (u) =  u^{-3/2} \delta\phi_{\omega,k} (u)$. The equation of motion to be solved then reads
\be   
\label{eq::eqQNM}
   4 u \left(u^6-1\right) \Phi ''(u) + 8 \left(2 u^6-2 i u^2 \bar\omega +1\right) \Phi '(u) + \left(9 u^5+8 i u\bar \omega + 16 u^3 \bar k^2
   \right) \Phi (u) = 0 \ .
\ee
Given the boundary conditions, only a discrete set of $\bar\omega$ give non-trivial solutions for each $\bar k$. 

To numerically determine these values, we adopt two paths. For the case of vanishing momenta, $\bar k = 0$, we adopt a standard shooting method. We take the expansion of the field at the boundary and at the horizon,
\begin{equation}
\label{eq::expansionsforQNM}
 \Phi_{\text{b}}(u) =  \sum _{n=0}^{\infty} j_{n} u^{n} \ , \quad \quad \quad  \Phi_{\text{h}}(u) = \sum _{n=0}^{\infty} h_{n} (1-u)^{n} \ .
\end{equation}
Notice that by expanding $ \Phi_{\text{h}}$ in this manner, we are selecting the ingoing solution, as the outgoing one is not regular at the horizon if one uses Eddington-Finklestein coordinates. Furthermore, the no-source boundary condition $\delta J=0$ establishes a relationship between $j_{3}$ and $j_{0}$ according to

\be
j_{3} = 3 \pr{\frac{4 \pi }{3}}^{-3/2} \pq{ \frac{a}{T^{3/2}} + \frac{2 b}{T^{3/4}}   \pr{\frac{\psi_{0}}{T^{3/4}}}  + 3 c  \pr{\frac{\psi_{0}}{T^{3/4}}} ^{2} } j_{0} \ ,
\ee
where $\psi_{0}$ is the expectation value of the equilibrium solution as defined in (\ref{eq::expvaluepsi}).
Solving the equations close to the boundary and the horizon, we find the coefficients $j_{n}$ and $h_{n}$. The first ones read
\ba
j_{1} = 0 \ , \quad \quad \quad j_{2} = - i \bar\omega j_{0} \ , 
\ea
and
\ba
h_{1} = - \frac{ 8\bar \omega -9 i}{8 (2 \bar\omega +3 i)} h_0 \ , \quad \quad h_{2} = -\frac{ 64\bar \omega ^2+1152 i \bar\omega -135 }{256 \left(2 \bar \omega ^2+9 i \bar\omega -9\right)} h_0 \ .
\ea
We then solve equation (\ref{eq::eqQNM}) numerically twice, utilizing the Taylor expansions described above as boundary conditions. This yields two numerical solutions, $ \Phi_{\text{h}}(u)$ depending on the free coefficient $h_{0}$, and $ \Phi_{\text{b}}(u)$ depending on the free coefficient $j_{0}$. These two solutions smoothly match at an arbitrary location $u=u_{0}$ along the radial direction only for a discrete set of values of $\bar \omega$ that render the Wronskian of the solutions zero. Therefore, by imposing
\be
 \Phi_{\text{h}} (u_{0})  \Phi_{\text{b}}' (u_{0}) - \Phi_{\text{b}} (u_{0}) \Phi_{\text{h}}' (u_{0})=0 \ ,
\ee
we numerically find the quasinormal mode values of
\be
\bar\omega = \bar\omega_{R} - i \Gamma \ .
\ee

For the case of finite momenta, $\bar k \neq 0$, we found the quasinormal mode frequencies by solving \eqref{eq::eqQNM} as a generalized eigenvalue problem (see e.g. \cite{Jansen:2017oag}). In this procedure, we discretize \eqref{eq::eqQNM} by using a Chebyshev grid, leading to the matrix problem
\be
(M_0 + \bar\omega M_1) \Phi = 0 \ ,
\label{eq::Generalized_Eigenvaluve_Problem}
\ee
with the matrices being
\be
M_0 = 4u(u^6-1)D^2_u + 8(2u^6+1)D_u+(9u^5 + 16 u^3 \bar k^2)\mathds{1} , \quad M_1 = -16iu^2D_u+8iu\mathds{1} \ .
\ee
Here, $D_u$ and $\mathds{1}$ refer to the derivative matrix on the Chebyshev grid and the identity one respectively.

Once the problem has been written in this way, we solved the system \eqref{eq::Generalized_Eigenvaluve_Problem} using the \textbf{LinearSolve} routine in \textit{Mathematica}. We did this for two different number of grid points, what allows one to distinguish those modes that are physical from those that are spurious. In all our cases we used $N_u = 40$ and $60$ grid points, with 30 digits of precision, keeping those eigenmodes whose relative difference was below $10^{-10}$.

\begin{figure}[t!]
\begin{center}
{\includegraphics[width=0.48\textwidth]{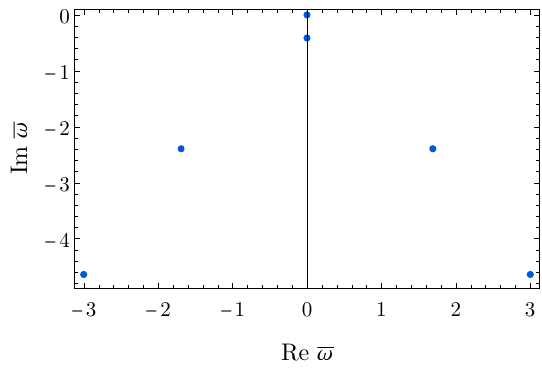}
\includegraphics[width=0.48\textwidth]{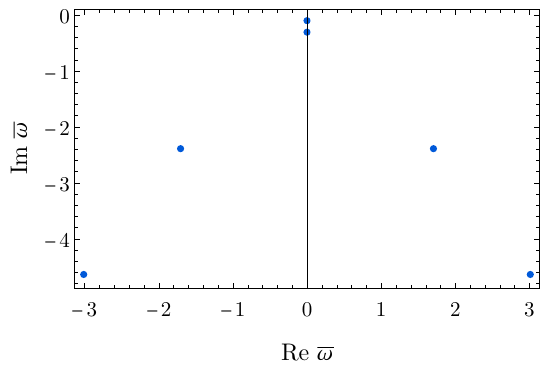}}
\end{center}
\caption{Lowest quasinormal mode frequencies for (left) $\bar k = 0$ and (right) $\bar k = 0.161$ and first-order phase transition in the first line of Table \ref{tab:parametVeff}. The chosen temperature is $T = T_2 + 10^{-8} T_c$.}
\label{fig::QNMs_example}
\end{figure}

An example of the result of the QNM computation is shown in Figure \ref{fig::QNMs_example}, for two different values of momenta, $\bar k = \{0, 0.161\}$, and first-order phase transition in the first row of Table \ref{tab:parametVeff}. The computation was done for a temperature very close to the spinodal point, $T= T_2 + 10^{-8}T_c$, so that the relaxation time of the scalar field is practically zero.

\section{Numerical methods for the time evolution}
\label{app::numericalcomputation}
In this appendix, we provide some details on the numerical approach that we have taken in order to study the time evolution of the scalar field in our holographic model. 

Assuming only time and radial dependence for the scalar field, $\phi=\phi(v,Z)$, the equation of motion we need to solve \eqref{eq:scalareom} becomes
\begin{equation}
\label{eq:scalareomHOM}
0= \partial_Z^2 \phi+\left(\frac{F'(Z)}{F(Z)}-\frac{2}{Z} \right)\partial_Z\phi+\frac{27/16}{Z^2 F(Z)}\phi +\frac{2}{Z F(Z)}\partial_v \phi-\frac{2}{F(Z)} \partial_Z\partial_v \phi \ .
\end{equation}
We introduce $\varphi_{-}$ and $\varphi_{+}$ as the dimensionless leading and subleading terms in the boundary expansion of the scalar field,\footnote{The relation between $\varphi_{\pm}$ and $\phi_{\pm}$ used in section \ref{sec:holographicModel} is
\begin{equation}
\phi_{\pm} = z_{h}^{- \Delta_{\pm}} \varphi_{\pm} = \pr{\frac{4\pi T}{3}}^{\Delta _{\pm}} \varphi _{\pm} \ .
\end{equation}}
\begin{equation}
    \phi(v,Z) = Z^{3/4}\left[ \varphi_-(v)+\partial_v\varphi_-(v)Z+\varphi_+(v)Z^{3/2}+\cdots\right] \ .
\end{equation}
Following the discussion in section \ref{subsec:effpotentialandmultitracebc}, we need to solve (\ref{eq:scalareom}) imposing regularity at the horizon $Z=1$ and the no-source, multitrace, boundary condition $J=0$, with $J$ given by (\ref{eq:J}). This can be written in terms of dimensionless quantities as follows
\begin{equation}
\label{eq:boundaryconditionappendix}
\varphi_++ \widetilde\Lambda \left(\frac{T}{T_{\text{first}}}\right)^{-9/4}-3  \widetilde a \left(\frac{ T}{T_{\text{first}}}\right)^{-3/2}\varphi_- +9  \widetilde b\left(\frac{ T}{T_{\text{first}}}\right)^{-3/4}\varphi_-^2-27c \varphi_-^3 =0\ .
\end{equation}
Here we have introduced the dimensionless coefficients 
\begin{equation}
   \widetilde \Lambda=\left(\frac{4\pi T_{\text{first}}}{3}\right)^{-9/4} \Lambda,\quad \widetilde a=\left(\frac{4\pi T_{\text{first}}}{3}\right)^{-3/2}a,\quad \widetilde b=\left(\frac{4\pi T_{\text{first}}}{3}\right)^{-3/4}b, 
\end{equation}
with the parameters $\{a, b, c, \Lambda \}$ chosen as in Table \ref{tab:parametVeff} depending on the case at hand. $T$ is the temperature evolving in time according to \eqref{eq:tempevol} and $T_{\text{first}}\approx 4.934 M$ is the critical temperature of the first-order phase transition in the first row of the table.

From the numerical point of view, it is more convenient to work with expansions displaying integer powers, so we change the holographic variable according to $Z = u^2$. Moreover, since we need to impose the multitrace boundary condition (\ref{eq:boundaryconditionappendix}) that relates the subleading and the leading terms in the field expansion, it turns out to be useful to formulate the problem introducing the field $\chi(v,u)$ through
\be
\phi (v,u) = u^{3/2} \pq{ \varphi_{-} (v) +  u^{2}  \p_{v} \varphi_{-} (v)+ u^{3} \chi (v,u)} \ .
\ee
The final equation of motion is,
\ba
\label{eq:finaleqtosolve}
 &&81 u^5 \chi  - 4 (4-10 u^{6}) \p_{u} \chi +4 u \pq{ 4 u \p_{u} \p_{v} \chi  - (1-u^{6}) \p_{u}^{2} \chi  +10 \p_{v} \chi}
   + \nb \\
   && +\ 49 u^4 \p_{v} \varphi_{-}  + 9 u^2  \varphi_{-}  + 24   \p_{v}^{2} \varphi_{-} = 0 \ .
\ea
whereas the boundary condition to be imposed is $\chi (0,v) = \varphi_{+} (v)$ with $\varphi_{+} (v)$ given by (\ref{eq:boundaryconditionappendix}).

We decided to numerically solve the problem by using the ``method of lines''. That is, we discretized the holographic direction $u$, including the horizon $u=1$ and the boundary $u=0$ as points of the discretization grid, transforming (\ref{eq:finaleqtosolve}) into a set of coupled ordinary differential equations for the function values at each grid point, $\chi_i(v)=\chi(u_i,v)$, with $i=1,2...N$, where $N$ is the number of grid points. By replacing $\partial_u$ with the numerical derivative matrix in \eqref{eq:finaleqtosolve}, we obtained the aforementioned system of equations.

We solved this system of ordinary differential equations for the variables $\varphi_- (v)$ and $\chi_i (v)$ utilizing the \textbf{NDSolve} routine of \textit{Mathematica}, choosing two different differentiation matrices: finite-differences and Chebyshev. We found that the results coming from the two choices agree. 

\bibliographystyle{JHEP}
\bibliography{Bibliography}

\end{document}